\documentclass[conference]{IEEEtran}
\usepackage{cite}
\usepackage{graphicx}

\ifCLASSOPTIONcompsoc
\usepackage[caption=false, font=normalsize, labelfont=sf, textfont=sf]{subfig}

\else
\usepackage[caption=false, font=footnotesize]{subfig}
\fi

\usepackage{amssymb}
\usepackage{amsmath}
\newtheorem{myDef}{Definition}
\newtheorem{mytheorem}{Theorem}
\newtheorem{myproof}{Proof}

\newtheorem{myproposition}{Proposition}

\usepackage[noend]{algpseudocode}
\usepackage{algorithmicx,algorithm}
\renewcommand{\algorithmicrequire}{ \textbf{Input}}      %Use Input in the format of Algorithm
\renewcommand{\algorithmicensure}{ \textbf{Output}}     %UseOutput in the format of Algorithm
\usepackage{array}
\usepackage{url}
\usepackage{multirow}
\usepackage{latexsym}
\usepackage{color}

\usepackage{amsmath}

\begin{document}
\title{When Homomorphic Cryptosystem Meets Differential Privacy: Training Machine Learning Classifier with Privacy Protection}

\author{\IEEEauthorblockN{Xiangyun Tang}
\IEEEauthorblockA{Beijing Institute of Technology\\
xiangyunt@bit.edu.cn}
\and
\IEEEauthorblockN{Liehuang Zhu}
\IEEEauthorblockA{Beijing Institute of Technology\\
liehuangz@bit.edu.cn}
\and
\IEEEauthorblockN{Meng Shen}
\IEEEauthorblockA{Beijing Institute of Technology\\
shenmeng@bit.edu.cn}
\and
\IEEEauthorblockN{Xiaojiang Du}
\IEEEauthorblockA{Temple University\\
dxj@ieee.org}}

\maketitle

\begin{abstract}
%\boldmath
Machine learning (ML) classifiers are invaluable building blocks that have been used in many fields.
High quality training dataset collected from multiple data providers is essential to train accurate classifiers.
However, it raises concern about data privacy due to potential leakage of sensitive information in training dataset.
Existing studies have proposed many solutions to privacy-preserving training of ML classifiers,
but it remains a challenging task to strike a balance among accuracy, computational efficiency, and security.

In this paper, we propose \texttt{Heda}, an efficient privacy-preserving scheme for training ML classifiers.
By combining homomorphic cryptosystem (HC) with differential privacy (DP), \texttt{Heda} obtains the tradeoffs between efficiency and accuracy, and enables flexible switch among different tradeoffs by parameter tuning.
In order to make such combination efficient and feasible,
we present novel designs based on both HC and DP:
A library of building blocks based on partially HC are proposed to construct complex training algorithms without introducing a trusted third-party or computational relaxation;
A set of theoretical methods are proposed to determine appropriate privacy budget and to reduce sensitivity.
Security analysis
demonstrates that our solution can construct complex
ML training algorithm securely.
Extensive experimental results
show the effectiveness and efficiency of the proposed scheme.
\end{abstract}


% \begin{center} \bfseries EDI$\mathcal{U}$ Category: 3-BBND \end{center}
% \fi
%%\IEEEpeerreviewmaketitle

\section{Introduction}\label{sec:introduction}
Machine learning (ML) classifiers are widely used in many fields, such as spam detection, image classification, and natural language processing.
Many studies have modeled user data and obtained satisfactory classifiers that meet accuracy requirements \cite{54,55}.
The accuracy of a classifier obtained from supervised learning is closely related to the quality of the training dataset, in addition to well-designed ML algorithms.
An experimental study with a dataset of 300 million images at Google \cite{61} demonstrates that the performance of classifiers increases as the order of magnitude of training data grows.
However, training dataset is usually held by multiple data providers and may contain sensitive information,
so it is important to protect data privacy in training of ML classifiers.

\begin{figure}[t]
\centering
\includegraphics[width=9.2cm]{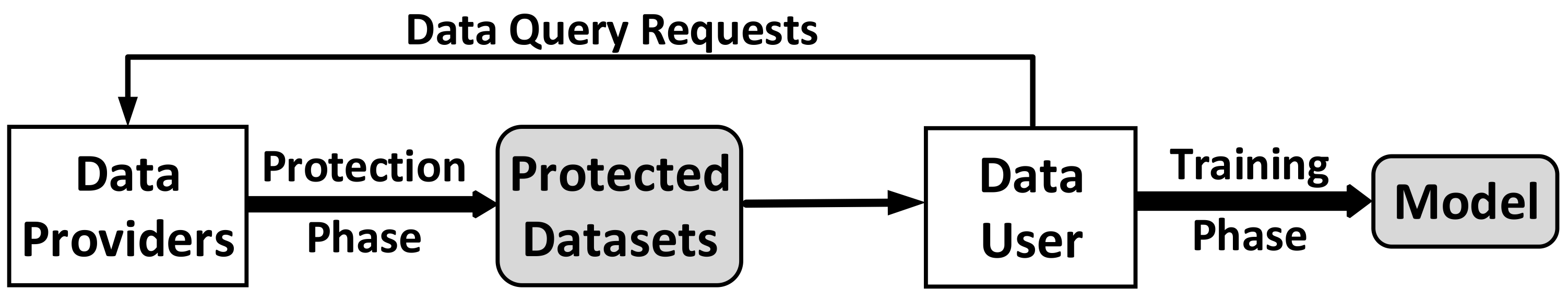}
\caption{Application Scenario.
Each non-shaded rectangle represents a type of role.
Each shaded box indicates private data that should be accessible to only one party: a protected dataset to a data provider, and the model to the data user.
Each solid arrow indicates an algorithm or a process.
%Single arrows indicate data flow.
}\label{fig:Application Scenario}
\end{figure}

Consider the typical training process depicted in Figure \ref{fig:Application Scenario}.
There are multiple data providers and a single data user.
Upon receiving the request of dataset from the data user,
each data provider applies privacy-preserving mechanisms (e.g., encryption or perturbation) to its own dataset.
Then, the data user trains an ML classifier based on the dataset gathered from multiple data providers.
During this process, each data provider cannot know the classifier,
while the data user cannot learn any sensitive information of the shared data.

More specifically, consider the following example of an advertisement recommendation task:
In order to attract more consumers, an company wants to build a classifier to discern the most appropriate time for advertising.
The training dataset used for constructing the classifier is extracted from the consumer purchase behavior data recorded by several online shopping sites.
The consumer data is confidential because it contains sensitive information about consumers.
Online shopping sites agree to share their data with companies, but refuse to reveal any privacy of the consumers.
The company wants to construct a classifier based on the consumer data, but is unwilling to reveal the classifier to online shopping sites.
Ideally, online shopping sites and the company run a privacy-preserving training algorithm, at the end of which the company learns the classifier parameters, and neither party learns anything else about the other party's input.

In general, supervised ML classifiers consist of two phases: the training phase and the classification phase.
A series of secure schemes for the classification phase have been proposed \cite{2,20}.
In this paper, we focus on the training phase, that is, privacy-preserving training of ML classifiers\footnote{In this paper, ML classifers and ML models are used interchangeably.}.

Existing solutions to training ML classifier securely roughly depend on three types of techniques, namely secure multi-party computing (SMC), homomorphic cryptosystem (HC), and differential privacy (DP).
SMC can construct many classifiers theoretically.
But it relies on a trusted third-party for providing random number, and results in a large number of interactions and redundant computations for protecting data privacy \cite{22,23}.
HC\footnote{In this paper, we only consider partially HC due to the computational inefficiency of fully HC.} allows the operation on ciphertext to be mapped to the corresponding plaintext.
The secure training solutions based on HC \cite{8,9}
may suffer from low efficiency.
In addition, since partially HC only enables a single type of operation (e.g., addition or multiplication),
HC-based solutions for training complex ML classifiers usually introduce a trusted third-party (e.g., the authorization server \cite{8,20}) or use an approximate equation that simplifies the complex iteration formula \cite{36,46}.
DP can resist the attacker with the largest background knowledge \cite{37}, which ensures the security of the published data by adding noises.
The computational efficiency of operations on perturbed data is significantly higher than those on ciphertext \cite{37,39}.
Nevertheless, the quality of the published dataset is reduced due to the introduction of noises, and thereby the accuracy of the resulting classifiers is decreased inevitably.

As discussed above, HC is low efficient due to ciphertext-based computation, but can obtain a classifier with lossless accuracy.
DP has high computational efficiency but leads to an inevitable loss of accuracy.
Intuitively, we can take the strengths of HC and DP by adopting them simultaneously.
%It is straightforward to a we can play to the strengths of DP and HC embedding DP into HC to get a high-efficiency and high-accuracy privacy-preserving ML training algorithm.

However, HC and DP are completely different systems: one for data encryption, and the other for data perturbation.
It is a challenging task to combine them together.
In particular,
partially HC only supports one type of operation,
which sets a barrier to the training of ML classifiers with complex operations such as power function, division, and square root.
Furthermore,
the noises added to sensitive data in DP determines the accuracy of classifiers and privacy of published data.
%reduces the quality of the training dataset and results in an inevitably reduced accuracy of the trained classifier.
%A high privacy budget or a low sensitivity may decrease the added noise.
%But if privacy budget is selected as a too large value, although some systems have been built upon DP framework, they doesn't mean that privacy is actually enforced by the system,
%so the privacy budget cannot be assigned arbitrarily.
%The sensitivity is only determined by the query function.
The third challenge is how to archive high accuracy while ensuring privacy in DP.

In this paper, we propose \texttt{Heda}, an efficient privacy-preserving scheme for training ML classifiers.
By combining HC with DP, \texttt{Heda} obtains the tradeoffs between efficiency and accuracy and enables flexible switch among different tradeoffs by parameter tuning.
%\texttt{Heda} is also flexible, i.e., one can determine the balance between the efficiency and accuracy of the algorithm by setting a parameter.
Security analysis demonstrates that our building blocks can construct complex ML training algorithms.
Extensive experimental results show the effectiveness and efficiency of the proposed scheme.

We address the above challenges by developing a set of key techniques.

We make an observation that different features\footnote{Without loss of generality, when facing the same training task, we assume that all the dataset has been locally preprocessed and represented with the same feature vectors \cite{7,8}.} in a dataset usually contribute differently to the accuracy of classifiers \cite{48,11}.
%Some features are crucial in training to construct accurate classifiers, while others are of less significance.
For the features with high contributions, we apply HC to these features such that the model parameters obtained from them are as accurate as possible.
We apply DP to the rest features to improve the computational efficiency.
The contribution of each feature in training ML classifiers can be evaluated using readily available techniques \cite{48}.
%Feature evaluation techniques determine whether the feature in the dataset is irrelevant or noisy to the analytical task by evaluating and scoring the contribution of each feature to the training of a model \cite{48}.
%
%Therefore, with feature evaluation techniques, we address the first challenge.
%(Section \ref{sec:combine})

%\vspace{-3mm}
To address the second challenge,
we employ two homomorphic encryption primitives: a multiplicative homomorphic encryption RSA and an additively homomorphic encryption Paillier.
We carefully design a library of building blocks supporting for complex operations such as power function and dot product, which can handle ML classifiers with complex training operations.
%Our HC based secure training algorithms need no the Authorization Server or any approximate equation to simplify the original iterative formula.
We take Logical Regression (LR) as an example to illustrate the power of our building blocks.
The sigmoid function in the iterative formula of LR makes it difficult to construct a secure LR training algorithm based on HC.
It is the first time that constructing a secure LR training algorithm by HC without an authorization server or any approximation.
(Section \ref{sec:HC})

%\vspace{-1mm}
In the face of the third challenge, we develop a formal method  to determine the reasonable privacy budget, and we reduce the sensitivity by using insensitive microaggregation.
We reduce the added noise and improve the usability of the noise dataset published by DP reasonably.
(Section \ref{sec:DP})

To the best of our knowledge, it is the first study that achieves privacy-preserving training of ML classifiers by jointly applying HC and DP in an individual scheme.
The rest of our paper is organized as follows.
Section \ref{sec:RW} describes related work,
and Section \ref{sec:Priliminaries} provides the background.
Section \ref{sec:Problem Statement} describes the problem statement.
Section \ref{sec:DP} and Section \ref{sec:HC} present the special designs with DP and HC, respectively.
%secure features scoring is described in Sections \ref{sec:Secure Scoring Features},
Section \ref{sec:combine} describes the construction of \texttt{Heda} in detail.
The security analysis is exhibited in Section \ref{sec:Security Analysis}, and the evaluation results are provided in Section \ref{sec:PE}.
Section \ref{sec:conclusion} concludes this paper.

\section{Related Work}\label{sec:RW}
Since our work is related to secure ML classifiers algorithms which can be broadly divided into two categories:
privacy-preserving classification and privacy-preserving training.
We give the literature review of both subjects.
Because \texttt{Heda} jointly applying HC and DP,
and there are some studies about combining HC with DP but not about secure classifiers training,
we present a discussion about these works.
We give an analysis of our novelty at last.

\subsection{Privacy-Preserving ML Classification}\label{sec:RW-Privacy-Preserving ML Classification}
A series of techniques have been developed for privacy-preserving ML Classification.
Wang et al. \cite{58} proposed an encrypted image classification algorithm based on multi-layer extreme learning machine that is able to directly classify encrypted images without decryption.
They assumed the classifier had been trained, and the classifier not confidential.
Grapel et al. \cite{26} constructed several secure classification algorithms by HC, while the parameters of trained classifiers are not confidential for classifiers users.
Zhu et al. \cite{59} proposed a secure nonlinear kernel SVM classification algorithm, which is able to keep users' health information and healthcare provider's prediction model confidential.

Several works have designed general (non-application specific) privacy-preserving protocols and explored a set of common classifiers by HC \cite{2,20}.
%These works  and building blocks to construct classifiers.
Usually, classification algorithms are simpler than training algorithms,
building blocks that are able to build classification algorithms can be powerless for complex training algorithms.

\subsection{Privacy-Preserving ML Classifier Training}\label{sec:RW-Privacy-Preserving ML Classifier Training}
Three techniques have been applied to privacy-preserving ML classifier training, they are SMC, HC, and DP.
%Existing solutions to training ML classifier securely roughly depend on three types of techniques,
%Secure two-party computations has been defined in  \cite{19},
%and it has been extended to SMC by Goldreich et al. \cite{22}.
%Lindell et al. has extended it to scenarios with active adversaries \cite{23}.
Constructing secure classifier training algorithms based on SMC relies on a large number of interactions and many redundant calculations for protect privacy,
and it generally needs to introduce authoritative third parties to provide random number distribution services as well.
In addition, SMC protocols for generic functions existing in practice rely on heavy cryptographic machinery.
Applying them directly to model training algorithms would be inefficient \cite{23,24,100}.

%In order to obtain efficient and secure compute protocols in a multi-party setting, HC have been studied \cite{2,26}.
HC is able to compute using only encrypted values.
Employing HC, many secure algorithms have been developed for different specialized ML training algorithms such as Support Vector Machine (SVM) \cite{8,28}, LR \cite{9,26}, decision trees \cite{33} and Naive Bayes \cite{30}.
However, partially HC only enables a single type of operation (e.g., addition or multiplication).
In order to construct complex training algorithms, HC-based schemes usually need to rely on trusted third parties such as the Authorization Server \cite{8,20}, or use an approximate equation to simplify the original complex iteration formula into a simple one \cite{36,46}.
Gonzlez et al. \cite{8} developed secure addition protocol and secure substractions protocol to construct the secure SVM training algorithm by employing Paillier,
while some operations that are not supported by Paillier have to be implemented with the assistance of the Authorization Server in their scheme.
Secure LR training algorithms existing implemented by HC are actually the linear regression \cite{9,26},
because the sigmoid function contains power function and division operation, which makes LR training algorithms harder to be implemented by HC than other ML training algorithms.
%because the sigmoid function makes it is complex to be computed than other ML training algorithm.
Several works solved the sigmoid function by an approximate equation\footnote{$\log (\frac{1}{1+\exp (u)})\approx \sum\limits_{j=0}^{k}{a\cdot u}$} \cite{36,46}.

Many secure ML classifier training algorithms have been explored in DP area such as decision tree \cite{37}, LR \cite{39} and deep learning \cite{40}.
Blum et al. \cite{37} proposed the first DP based decision tree training algorithm on the SuLQ platform.
%Rana et al. proposed a practical approach to ensemble decision trees in a random forest \cite{38}.
Abadi et al. \cite{40} applied DP objective perturbation in a deep learning algorithm, where the noise was added to every step of the stochastic gradient descent.
Due to the introduction of noise under DP mechanisms, the quality of the datasets were reduced, and the accuracy of these trained models was decreased inevitably.
So the essential challenge for DP based frameworks is guaranteeing the accuracy by reducing the added noise, especially for the operation has high sensitivities \cite{4}.
%How to reasonably reduce the noise added while meeting the privacy requirements is the key to applying DP to ML.
According to the Laplace mechanism (cf. Definition \ref{def:Laplace mechanism}), privacy budget $\epsilon $ and the sensitivity $\Delta f$ are two important factors affecting noise addition.
In many papers, the value of $\epsilon $ is merely chosen arbitrarily or assumed to be given \cite{37,40}.
%if $\epsilon $ is selected as a too large value, although some systems have been built upon DP framework, they doesn't mean that privacy is actually enforced by the system.
%Hsu et al. \cite{42} examined the role of $\epsilon $ in concrete applications.
Lee et al. \cite{16} explored the selection rules of $\epsilon $ , but they have not given a way to determine the value of the privacy budget.
%In addition to $\epsilon $, $\Delta f$ is also the other important parameter that determines the addition of noise.
Soria et al. \cite{17} proposed a insensitive microaggregation-based DP mechanism, they found the amount of noise required to fulfill $\epsilon $-DP can be reduced in insensitive microaggregation.
\texttt{Heda} develops the insensitive microaggregation-based DP mechanism and decreases the amount of noise required to fulfill $\epsilon $-DP again.

\subsection{Homomorphic Cryptosystem Combine Differential Privacy}\label{sec:RW-Homomorphic Cryptosystem Combines with Differential Privacy}
Several works have studied combining HC with DP to solve a special security problem.
Pathak et al. \cite{44} proposed a scheme for composing a DP aggregate classifier using classifiers trained locally by separate mutually untrusting
parties,
where HC was used for composing the trained classifiers.
Yilmaz et al. \cite{45} proposed a scheme for optimal location selection utilizing HC as the building block and employing DP to formalize privacy in statistical databases.
Aono et al. \cite{46} constructed a secure LR training algorithm via HC and achieved DP to protect the model parameters.
%In this paper, we focus on the ML training phase and improving the efficiency and the accuracy of the training algorithms.
%These works general constructed a secure algorithm via HC and used differential privacy to protect the results.
These works general constructed a secure algorithm via HC and used DP to protect the algorithm results.
As we have discussed above,
constructing a secure algorithm via HC is low efficient,
and secure algorithm based on DP has inevitable loss in accuracy.
We aim of constructing a secure classifier training algorithm jointly applying HC and DP in an individual scheme to obtain a tradeoff between efficiency and accuracy.

\subsection{Novelty of Our Construction}\label{sec:RW-Novelty}
Secure training algorithms based on HC have to handle datasets in ciphertext case, where the time consumption is considerable,
while the accuracy is able to be guaranteed.
Noise datasets published by DP mechanism are in plaintext case, it is efficient to train a model in plaintext case, while using the noise dataset may lead to a low accuracy.
HC and DP have drawbacks as well as merits.

\texttt{Heda} takes the strengths of HC and DP to get a high-efficiency and high-accuracy privacy-preserving ML classifier training algorithm.
\texttt{Heda} is the first to combine these two techniques and construct a privacy-preserving ML classifier training algorithm in a multi-party setting, where feature evaluation techniques are employed to give the way of combination.
By combining HC with DP, \texttt{Heda} obtains the tradeoffs between efficiency and accuracy, and enables flexible switch among different tradeoffs by parameter tuning.
%\texttt{Heda} is able to balance accuracy and efficency through the combination of HC and DP.
What's more,
we develop a library of building blocks by HC that is able to construct complex training algorithms,
and by using our building blocks this is the first time that solving the sigmoid function in secure LR training based on HC without any approximate equation.
We develop the works of Lee et al. \cite{16} and Soria et al. \cite{17} giving a formula to determine the appropriate privacy budget and another lower sensitive solution.

\section{Priliminaries}\label{sec:Priliminaries}
\subsection{Notation}\label{sec:Priliminaries-Notation}
A dataset $D$ is an unordered set of n records with the size of $|D|$.
${{x}_{i}}\in {{\mathbb{R}}^{d}}$, ${x}_{i}=\left( {{x}_{i1}},{{x}_{i2}},\ldots,{{x}_{id}} \right)$ is the i-th record in dataset $D$, and ${{y}_{i}}$ is a class label correspond to ${{x}_{i}}$.
$X = \left( {{x}_{1}},{{x}_{2}},\ldots,{{x}_{m}} \right)$, $Y = \left( {{y}_{1}},{{y}_{2}},\ldots,{{y}_{m}} \right)$.
$\beta,\omega$, and $b$ are the relevant parameters of the model trained by a ML algorithm.
The subset $A_i$ corresponding to the i-th attribute in $D$.
S is the scores assign to the features.
% of several data providers after feature evaluation.
\\\indent
Cryptosystems define a plaintext space $\mathbb{M}$, and a ciphertext space $\mathbb{C}$.
%$\phi \left( N \right)$ is the Euler phi-function\footnote{Let $N>1$ be an integer.
%Then $Z_{N}^{*}$ is an abelian group under multiplication modulo N.
%Define $\phi \left( N \right)\underline{\underline{def}}\left| Z_{N}^{*} \right|$, the order of the group $Z_{N}^{*}$.}, and the group ${{\mathbb{Z}}_{N}}$ has the set $\left\{ 0,...,N-1 \right\}$ respect to addition modulo N.
In \texttt{Heda}, we employ two public-key cryptosystems, Paillier and RSA.
$[[ m ]]$ and $||m||$ are represented as the ciphertext of m under Paillier or RSA respectively.
\\\indent
DP is generally achieved by a randomized algorithm $\mathcal{M}$.
$\epsilon $ is the privacy budget in a DP mechanism.
A query $f$ maps dataset D to an abstract range $f:D\to R$.
The maximal difference in the results of query $f$ is defined as the sensitivity $\Delta f$.
$D'$ is a neighboring dataset of $D$.
\\\indent
Table \ref{table:Notations} summarizes the notations used in the following sections.

\begin{table}[!t]
\normalsize
\centering
\small
\caption{Notations}\label{table:Notations}
\renewcommand{\arraystretch}{1.1}
\begin{tabular}{p{0.7cm}p{2.9cm}p{.7cm}p{2.9cm}}
\hline
Notations	&\ \ \ \ \ Explanation	&Notations	&\ \ \ \ \ Explanation
\\
\hline
\hline
$\mathbb{R}$	&Set of real numbers	&	${{\mathbb{R}}^{d}}$	&	d-dimension $\mathbb{R}$
\\
$D$		&Dataset	&	$|D|$	&	The size of D
\\
$m$	&	Size of dataset	&	$D'$		&Neighbour dataset
\\
$X$		&The record set in D		&	$Y$		&The label set in D
\\
${{x}_{i}}$		&i-th Record in dataset		&	${{y}_{i}}$		&Class label
\\
$o,\sigma$		&Functional operation		&	  $d$	&	Dataset dimension
\\
$\mathbb{M}$		&Plaintext space			&$\beta, b$		&Parameters of models
\\
$\mathcal{M}$		&Mechanism	&	$\mathbb{C}$	&	Ciphertext space
\\
$k$		&The cluster size	&	$N$	&	  n-bit Primes
\\
$f$	&	Query	&		$\epsilon $	&	Privacy budget
\\
$\gamma$	&	Noise		&		$\Delta f$&	Sensitivity
\\
$\left[\left[ m \right]\right]$		&Ciphertext under Paillier		&	$||m||$		&Ciphertext under RSA
\\
$\iota$ & The number of encrypted features in D&S &The scores of features
\\
$[m]$		& The encryption of m under a certain cryptosystems	&		${A}_{i}$&The subset of i-th attribute in D
 \\
\hline
\end{tabular}
\end{table}

\subsection{Homomorphic Cryptosystem}\label{sec:Priliminaries-HC}
Cryptosystems are composed of three algorithms: key generation (Gen) to generate the key, encryption (Enc) encrypting secret message and decryption (Dec) for decrypting ciphertext.
Public-key cryptosystems employ a pair of keys ($\sf{PK}$, $\sf{SK}$), the public key ($\sf{PK}$, the encryption key) and the private key ($\sf{SK}$, the decryption key).
Some cryptosystems are gifted with a property of homomorphic that makes cryptosystems perform a set of operations on encrypted data without knowledge of the decryption key.
Formalized definition is given in Definition \ref{def:homomorphic}.
\begin{myDef}\label{def:homomorphic}(homomorphic) \cite{1}.
 A public-key encryption scheme $\left( \text{Gen, Enc, Dec} \right)$ is homomorphic if for all $n$ and all $\left( \sf{PK}, \sf{SK} \right)$ output by $\text{Gen}\left( {{1}^{n}} \right)$, it is possible to define groups $\mathbb{M}$, $\mathbb{C}$ (depending on $\sf{PK}$ only) such that:
\\\indent
\emph{(i) The message space is $\mathbb{M}$, and all ciphertexts output by $\text{Enc}_{pk}$ are elements of $\mathbb{C}$.}
\\\indent
\emph{(ii) For any ${{m}_{1}},{{m}_{2}}\in \mathbb{M}$, any ${{c}_{1}}$ output by ${{Enc}_{pk}}\left( {{m}_{1}} \right)$, and any ${{c}_{2}}$ output by ${Enc}_{pk}\left( {{m}_{2}} \right)$, it holds that ${{Dec}_{sk}}\left( o\left( {{c}_{1}},{{c}_{2}} \right) \right)=\sigma \left( {{m}_{1}},{{m}_{1}} \right)$.}
\end{myDef}

In \texttt{Heda}, we employ two public-key cryptosystems, Paillier and RSA.
Paillier possesses additively homomorphic property, and RSA possesses multiplicative.
For more details about Paillier or RSA, we refer the reader to \cite{1}.

\textbf{Paillier.}
The security of Paillier is based on an assumption related to the hardness of factoring.
%Let $p$ and $q$ are n-bit primes, $N=pq$.
%The public key is $N$, and the private key is $\left( N, \phi (N)\right)$ in Paillier.
%Encryption function in paillier is $c:=\left[\left[ {{\left( 1+N \right)}^{m}}{{r}^{N}} mod {{N}^{2}} \right]\right]$, where $m\in {{\mathbb{Z}}_{N}}$.
%Decryption function in paillier is $m:=\left[\left[ \frac{\left[ {{c}^{\varnothing \left( N \right)}}mod{{N}^{2}}-1 \right]}{N}.\phi {{\left(N \right)}^{-1}}mod N \right]\right]$.
%For more details about paillier, we refer the reader to \cite{1}.
Assuming a pair of ciphertext $\left( {{c}_{1}},{{c}_{2}} \right)$ is $\left( {{m}_{1}},{{m}_{2}} \right)$ under the same Paillier encryption scheme where the public key is $N$, we have: ${{c}_{1}}\times {{c}_{1}}=\left[\left[ {{\left( 1+N \right)}^{{{m}_{1}}+{{m}_{2}}}}{{r}^{N}}mod{{N}^{2}} \right]\right]$, where $\left( {{m}_{1}}+{{m}_{2}} \right)<N$.
The additively homomorphic property in Paillier can be described as $\left[\left[ {{m}_{1}}+{{m}_{2}} \right]\right]=\left[\left[ {{m}_{1}} \right]\right]\times \left[\left[ {{m}_{2}} \right]\right]\left( mod{{N}^{2}} \right)$.

\textbf{RSA.}
%RSA was proposed in 1978 by R.L. Rivest, A. Shamir, and L. Adleman \cite{3}.
Based on the definition of a one-way trapdoor function, RSA gives the actual implementation of the first public key cryptosystem.
%Let $p$ and $q$ are n-bit primes, $N=pq$.
%$1\le \text{e}\le \phi \left( N \right), d ={{e}^{-1}}mod\phi \left( N \right)$.
%The public key of RSA is: (N, e), and private key is: (N, d).
%The RSA encryption function is: $c:= ||{{m}^{e}} modN||$, and the decryption function is: $m:= ||{{c}^{d}} modN ||$.
RSA is a multiplicative HC, because that: $En{{c}_{RSA}}\left( {{m}_{1}} \right)\times En{{c}_{RSA}}\left( {{m}_{2}} \right)=||{{\left( {{m}_{1}}\times {{m}_{2}} \right)}^{e}}modN||$, where $\left( {{m}_{1}}\times{{m}_{2}}\right)<N$.
The multiplicative homomorphic property in RSA can be described as $||{{m}_{1}}\times {{m}_{2}}||=||{{m}_{1}}||\times ||{{m}_{2}}||\left( modN \right)$.

\subsection{Differential Privacy}\label{sec:Priliminaries-DP}

\begin{myDef}\label{def:Neighbor Dataset}(Neighbor Dataset) \cite{37}.
The datasets $D$ and $D'$ have the same attribute structure, and the symmetry difference between them is denoted as $|D\vartriangle D'|$.
We call $D$ and $D'$ neighbour datasets if $\left| D\vartriangle {D}' \right|=1$.
\end{myDef}

\begin{myDef}\label{def:DP}($ \epsilon$-Differential Privacy) \cite{37}.
A randomized mechanism $\mathcal{M}$ gives $ \epsilon $-DP for every set of outputs $\mathcal{R}$, and for any neighbor dataset of $D$ and $D'$, if $\mathcal{M}$ satisfies:
		$\text{Pr}\left[ \mathcal{M}\left( D \right)\in \mathcal{R} \right]\le \exp \left( \epsilon  \right)\times \text{Pr}\left[ \mathcal{M}\left( D' \right)\in \mathcal{R} \right]$.
\end{myDef}

A smaller $\epsilon $ represents a stronger privacy level \cite{4}.
%Privacy level reaches its maximum when $\epsilon $ is equal to 0.
While $\epsilon$ is equal to 0, for any neighbour dataset, the randomized mechanism $\mathcal{M}$ will output two identical results of the same probability distribution which cannot reflect any useful information.
If $\epsilon $ is selected as a too large value in a DP mechanism, it does not mean that privacy is actually enforced by the mechanism.
%Therefore, the value of $\epsilon $ have to be carefully designed.
%$\epsilon $ controls the level of privacy guarantee achieved by mechanism.
A composition theorem for $\epsilon $ named parallel composition (Theorem \ref{theorem:Parallel Composition}) is widely used.

\begin{mytheorem}\label{theorem:Parallel Composition}(Parallel Composition) \cite{5}.
Suppose we have a set of privacy mechanisms $\mathcal{M}=\left\{ {{\mathcal{M}}_{1}},{{\mathcal{M}}_{2}},\ldots \ldots,{{\mathcal{M}}_{m}} \right\}$.
If each ${{\mathcal{M}}_{i}}$ provides a ${{\epsilon }_{i}}$-DP guaranteed on a disjointed subset of the entire dataset, $\mathcal{M}$ will provide $\left( max\left\{ {{\epsilon }_{1}},{{\epsilon }_{2}},\ldots \ldots,{{\epsilon }_{m}} \right\} \right)$-DP.
\end{mytheorem}

Lapace Mechanism (Definition \ref{def:Laplace mechanism}) is the basic DP implementation mechanism and is suitable for the numerical data, which adds independent noise following the Laplace distribution to the true answer.

\begin{myDef}\label{def:Laplace mechanism}(Laplace mechanism) \cite{60}.
For a dataset D and a query function $f:D\to R$ with sensitive $\Delta f$.
Privacy mechanisms $\mathcal{M}\left( D \right)=f\left( D \right)+\gamma$ providers $\epsilon $-DP, where $\gamma \sim lap \left( \frac{\Delta f}{\epsilon } \right)$ represents the noise sampled from a Laplace distribution with a scaling of $\left( \frac{\Delta f}{\epsilon } \right)$.
\end{myDef}

\begin{myDef}\label{def:Sensitivity}(Sensitivity) \cite{37}.
For a query $f:D\to R$, and a pair of neighbor datasets ($D$, $D'$), the sensitivity of $f$ is defined as: $\Delta f=\underset{D,D'}{\mathop{\max }}\,{{\left| \left| f\left( D \right)-f\left( D' \right) \right| \right|}_{1}}$.	
Sensitivity $\Delta f$ is only related to the type of query $f$.
It considers the maximal difference between the query results.
\end{myDef}

\section{Problem Statement}\label{sec:Problem Statement}
We are devoted to addressing the problem on the secure training of ML classifier using private protected data gathered from different data providers.
In this section, we introduce the overview of the system model and the roles involved in \texttt{Heda}.
Then, we formally define the threat model and the security goal.

\subsection{System Model}\label{sec:PS-System Model}
We target at the system application scenario which has been illustrated in Figure \ref{fig:Application Scenario}.
%Data providers provide data for data user, so service providers also named data providers.
There are $n$ data providers $\mathcal{P}$ and a data user $\mathcal{U}$ in our model.
Each $\mathcal{P}$ holds their own sensitive dataset $D_{i}$ and a pair of keys ($\sf{PK}$, $\sf{SK}$).
%Without loss of generality, when facing the same training task, we assume that all the data has been locally preprocessed and represented with the same feature vectors \cite{7,8}.
$\mathcal{P}$ protects their sensitive data by applying privacy-preserving mechanisms (e.g., DP mechanism and HC).
$\mathcal{U}$ holds his own keys ($\sf{PK}$, $\sf{SK}$).
After obtaining the permission, $\mathcal{U}$ requests the sensitive data from $\mathcal{P}$, and $\mathcal{P}$ returns the protected data.
%$\mathcal{U}$ interacts with $\mathcal{P}$ and receives receiving encrypted data from $\mathcal{P}$.
By running a sequence of secure interactive protocols with several $\mathcal{P}$,
$\mathcal{U}$ obtains the classifier parameters of being encrypted by $\mathcal{U}$'s keys.

As discussed in Section \ref{sec:introduction} and \ref{sec:RW},
HC is able to construct accurate secure training algorithms, and DP mechanism providers high efficient secure training algorithms.
However, it is low efficient that constructing a secure ML training algorithm by HC, and the model may poor in accuracy if the training data is under DP mechanism.
We thereby desire to take the strengths of HC and DP, and feature evaluation techniques is used for providing a right combination method.
%The construction of \texttt{Heda} model is exhibit in Figure \ref{fig:SystemModel},
We describe the overall idea of \texttt{Heda} as follows:
\\\indent
1) $\mathcal{P}$ scores all features by feature evaluation techniques and divides the dataset into two parts according to the scores
(see Section \ref{sec:combine-Dividing a Aataset into Two Parts According to the Feature Scores}).
\\\indent
2) $\mathcal{P}$ applies privacy-preserving mechanisms to the two parts respectively:
the low scores part published by DP mechanism (see Section \ref{sec:DP});
the high scores part encrypted by HC (see Section \ref{sec:HC}).
\\\indent
3) Upon receiving the query requests, $\mathcal{P}$ sends the protected data to $\mathcal{U}$.
\\\indent
4) $\mathcal{U}$ trains a ML classifier under these two protected sub-datasets (see Section \ref{sec:combine-Combining DP Mechanism with Our Building Blocks}).
\\\indent
%We will detail  confidentially in Section \ref{sec:Secure Scoring Features}.
%The DP and HC parts of \texttt{Heda} for Encryption will be presented in Section \ref{sec:DP} and \ref{sec:HC}.
%Section \ref{sec:combine} will describe the details of how to perform feature scoring and the combination of the two parts for training.

\subsection{Threat Model}\label{sec:PS-Threat Model}
$\mathcal{U}$ interacts with several $\mathcal{P}$ to obtain the protected data and performs training algorithms on the data.
Each $\mathcal{P}$ trys to learn as much other $\mathcal{P}$'s sensitive data and $\mathcal{U}$'s trained classifier as possible by honestly executing pre-defined protocols.
$\mathcal{U}$ follows the protocol honestly, but it tries to infer $\mathcal{P}$'s sensitive data as much as possible from the values he learns.
As discussed above, we assume each participant is a passive (or honest-but-curious) adversary \cite{10}, that is,
it does follow the protocols but tries to infer others' privacy as much as possible from the values they learn.

\subsection{Security Goal}\label{sec:PS-Security Goal}
In \texttt{Heda}, we allow any two or more parties conspire to steal the privacy of other participants.
We make the following assumptions:
Each participate as a honest-but-curious adversary performs protocol honestly but may have interest in the private information of other domains.
Any two or more participates may collude with each other.
As passive adversaries, they do follow the protocol but try to infer other's privacy as much as possible from the values they learn.

The aim of \texttt{Heda} is achieving keeping privacy of each participant and computing model parameters securely when facing honest-but-curious adversaries or any collusion.
To be specific, the privacy of $\mathcal{U}$ is model parameters, and each $\mathcal{P}$ is their sensitive data.
We specify our security goals as follows:
\begin{enumerate}
\item When facing honest-but-curious adversaries, $\mathcal{U}$ and each $\mathcal{P}$'s privacy are confidential.
\item when facing any two or more parties collude with each other, $\mathcal{U}$ and each $\mathcal{P}$'s privacy are confidential.
\end{enumerate}

\section{Accuracy and Privacy Design with Differential Privacy}\label{sec:DP}
DP ensures the security of the published data by adding noise.
Insufficient noise leads to the security of the published data cannot be guaranteed,
while excess noise causes the data unusable.
Obviously, the key to using DP in the secure classifier training is to reduce the added noise while ensuring the security of the published data.

The two important parameters that determine the added noise are $\epsilon $ and $\Delta f$ (cf. Definition \ref{def:Laplace mechanism}).
A bigger $\epsilon $ or a smaller $\Delta f$ are able to reduce the added noise.
However,
if $\epsilon $ is selected as a too large value, although the system has been built upon DP framework, it dose not mean that privacy is actually enforced by the system.
Therefore, $\epsilon$ must be combined with specific requirements to achieve the balance of security and usability of output results.
On the other hand,
$\Delta f$ is only determined by the type of query function (cf. Definition \ref{def:Sensitivity}).

In this section, we develop a formula for reasonably determining the appropriate $\epsilon $ in DP mechanism,
and we reduce the $\Delta f$ by using insensitive microaggregation.

%data can be anonymized from single data to group data by clustering, and clustered dataset can reduce the $\Delta f$ of query functions.
%, thereby reducing the loss of data information and improving data usability.
%In this section, we describe our development in the part of DP, including how to choose a appropriate $\epsilon $ in DP mechanism and how to use insensitive microaggregation to reduce the sensitivity $\Delta f$,
%and formal security proofs are given along with it.

\subsection{Selection of Appropriate $\epsilon $} \label{sec:DP-Appropriate e}
In many papers, $\epsilon $ is chosen arbitrarily or assumed to be given,
while decision on $\epsilon $ should be made carefully with considerations of the domain and the acceptable ranges of risk of disclosure.
Lee et al. \cite{16} explored the rule of $\epsilon $,
but they did not give a specific method for determining $\epsilon $.
We give a method for determining $\epsilon $.
It is worth noting that based on different criteria and backgrounds, $\epsilon $ can have different values,
and we are trying to give a general one.

We follow some notations of Lee et al. \cite{16}:
%For the neighbor dataset $D'$ of dataset $D$,
If an adversary knows all the background knowledge, he tries to guess which one is the different values between $D'$ and $D$.
Let $\mathbb{W}$ denotes the set of all possible combinations $\omega $ of $D'$, $\omega \in \mathbb{W}$.
For each possible $\omega $, the adversary maintains a set of tuples $\left\langle  \alpha , \mu  \right\rangle $.
For a given query response, $ \alpha $ and $ \mu $ are the adversary's prior belief and posterior belief on ${D}'$, i.e., $\forall \omega \in \mathbb{W}, \alpha \left( \omega  \right)=\frac{1}{m}$.
For each possible $\omega $, the adversary's posterior belief on $\omega $ is defined as
$ \mu \left( \omega  \right)=P\left( {D}'=\omega  | \gamma  \right)=\frac{P\left( \mathcal{M}\left( \omega  \right)=\gamma  \right)}{\mathop{\sum }_{ \omega \in \mathbb{W}}P\left( \mathcal{M}\left(  \omega  \right)=\gamma  \right)}$.
Lee et al. \cite{16} obtain the upper bound of $\epsilon $ through a series of derivations as Formula \ref{equ:others-the upper bound of the parameter} (cf. Section V in \cite{16})
\begin{equation}\label{equ:others-the upper bound of the parameter}
\epsilon \le \frac{\Delta f}{\Delta v}ln\left( \frac{\left( m-1 \right)\rho }{1-\rho } \right)
\end{equation}
where ${\Delta v} ={{max_{1\le i,j\le n,i\ne j}}| f( {{\omega }_{i}} )- f( {{\omega }_{j}} ) |}$, $\rho $ is the probability that the adversary guessing success.
%We can observe from Formula \ref{equ:others-the upper bound of the parameter} that since $n$ is the size of the dataset, when $n$ is very large, $\epsilon $ takes a larger value,
Nevertheless, Lee et al. \cite{16} did not give a method for setting $\rho $.
We give a method for determining the upper bound of $\rho$ (Proposition \ref{proposition:the upper bound of p}).

\begin{myproposition}[the upper bound of $\rho$ for ${D}'$]\label{proposition:the upper bound of p}
Let ${{A}_{j}}$ is the subset of the j-th attribute in dataset $D$.
%, ${{x}_{ij}}\in {{A}_{i}},0\le j\le n$.
%${{x}_{ij}}$ as ${{Count}_{max}}$ is the record has the highest frequency in ${{A}_{i}}$.
${{Count}_{max}}$ is the occurrences number of the record which has the highest frequency in ${{A}_{j}}$.
Then $\frac{{Count}_{max}}{\left| {{A}_{j}} \right|}$ is the upper bound of $ \rho $.	
\end{myproposition}

\begin{myproof}[Proof of Proposition \ref{proposition:the upper bound of p}]\label{myproof:the upper bound of p}
$\rho $ is the probability that the adversary successfully guesses which instance is the different one between $D'$ and $D$.
DP mechanism assumes that the adversary has a strong background knowledge, that is, he knows the value of each instance in $D$.
${{x}_{ij}}$ is the highest frequency instance in ${{A}_{j}}$, so the adversary guesses ${{x}_{ij}}$ will get the highest probability of success.
After DP mechanism, the adversary's probability of success should not be greater than the highest probability of random guessing and success,
so the upper bound of $\rho $ is $\frac{Coun{{t}_{max}}}{\left| {{A}_{j}} \right|}$.$\hfill\square$
\end{myproof}

%The noise dataset is generated by DP noise response that is call for the values of all the attributes in each record.
The upper bound of ${{\epsilon }_{i}}$ is obtained form each subset ${{A}_{i}}$ by Formula \ref{equ:others-the upper bound of the parameter},
then the dataset $D$ provides $max\left( {{\epsilon }_{i}} \right)$-DP according to Theorem \ref{theorem:Parallel Composition}.
Algorithm \ref{algorithm:generating the appropriate value of e} details the steps for generating the appropriate $\epsilon$ on dataset D.

\begin{algorithm}[!t]
\renewcommand\baselinestretch{1.2}\selectfont
\small
\caption{Generating Appropriate Value of $\epsilon$}\label{algorithm:generating the appropriate value of e} % 算法的名字
\begin{algorithmic}[1]
\Require \textbf{:} $D=\left\{ \left( {{x}_{i}},{{y}_{i}} \right) \right\}_{i=1}^{m}$.
\Ensure \textbf{:} The appropriate $\epsilon $ on dataset $D$.
\For{$j=1$ to d}
   \For{$i=1$ to m}
     \State Computing $\Delta f$ and $\frac{Coun{{t}_{max}}}{\left| {{A}_{j}} \right|}$ in ${{A}_{j}}$.
   \EndFor
    \State Obtaining ${{\epsilon }_{j}}$ by Formula \ref{equ:others-the upper bound of the parameter}.
\EndFor\\
\Return $\epsilon =\left\{ {{\epsilon }_{1}},{{\epsilon }_{2}},\ldots \ldots ,{{\epsilon }_{d}} \right\}$.
\end{algorithmic}
\end{algorithm}

\subsection{Reducing $\Delta f$ by Insensitive Microaggregation}\label{sec:DP-Reduce f by microaggregation}
According to the Definition \ref{def:Laplace mechanism}, the smaller the $\Delta f$, the less noise is added, and thereby the more usable the data is.
In this subsection, we detail the solution of reducing the $\Delta f$ in \texttt{Heda}.
The amount of noise required to fulfill $\epsilon $-DP can be greatly reduced if the query is run on a insensitive microaggregation version of all attributes instead of running it on the raw input data \cite{17}.

\subsubsection{What is insensitive microaggregation}
Microaggregation is used to protect microdata releases and works by clustering groups of individuals and replacing them by the group centroid.
%Using microaggregation can reduce $\Delta f$, then the noise that needs to be added in data is reduced.
DP makes no assumptions about the adversary's background knowledge.
%The synergy between microaggregation and $\epsilon $-DP can achieve more accurate and general-purpose $\epsilon $-DP:
Microaggregation with DP can help increasing the utility of DP query outputs while making as few assumptions on the type of queries as microaggregation does \cite{17}.
However, if we modify one record in $D$, more than one clusters will differ from the original clusters generally.
According to the Definition \ref{def:DP}, we expect that we modify one record in $D$, each pair of corresponding clusters differs at most in single record.
Microaggregation that satisfies this property is named insensitive microaggregation (IMA).
Soria et al. \cite{17} give a formal definition of IMA.
Microaggregation is insensitive to input data if and only if the distance function $Dist(x,y)$ is a fixed sequence of total order relations defined over the domain of $D$ \cite{17}.
%\begin{myDef}\label{def:IMA}(IMA). \cite{17}
%Let $D$ be a dataset, $M$ a microaggregation algorithm, and let $\left\{ {{C}_{1}},\ldots ,{{C}_{n}} \right\}$ be the set of clusters that result from running $M$ on $D$.
%Let $D'$ be a data set that differs from $D$ in a single record, and $\left\{ C_{1}^{'},...,C_{n}^{'} \right\}$ be the clusters produced by running $M$ on $D'$.
%We say that $M$ is insensitive to the input data if, for every pair of datasets $D$ and $D'$ differing in a single record, there is a bijection between the set of clusters $\left\{ {{C}_{1}},\ldots ,{{C}_{n}} \right\}$ and the set of clusters $\left\{ C_{1}^{'},...,C_{n}^{'} \right\}$ such that each pair of corresponding clusters differs at most in a single record.
%\end{myDef}

The sequence of total orders is determined by a sequence of reference points.
The reference points are the two boundary points $P$ and $P'$,
i.e. $P=\left( {{p}_{1}},{{p}_{2}},\ldots ,{{p}_{d}} \right)$, ${{p}_{i}}=\max \left( {{A}_{i}} \right)$,
and $P' =\left( {p'}_{1},{p'}_{2},\ldots ,{p'}_{d} \right)$, $p_{i}^{'}=\min \left( {{A}_{i}} \right)$.
The total order relations between two points in \texttt{Heda} is:
$Dist\left( x,y \right)=\sqrt[{}]{\underset{i=0}{\overset{d}{\mathop \sum }}\,\frac{{{\left( {{x}_{i}}-{{y}_{i}} \right)}^{2}}}{{{\left( {{P}_{i}}-P_{i}^{'} \right)}^{2}}}}$.
Generating a IMA dataset is detailed in Algorithm \ref{algorithm:IMA}.

\begin{algorithm}[!t]
\renewcommand\baselinestretch{1.2}\selectfont
%\floatname{algorithm}{Protocol}
\small
\caption{Generating an IMA Dataset}\label{algorithm:IMA} % 算法的名字
\begin{algorithmic}[1]
\Require \textbf{:} $D=\left\{ \left( {{x}_{i}},{{y}_{i}} \right) \right\}_{i=1}^{m}$ ,k is the cluster size, $m\ge 2k$.
\Ensure \textbf{:} A IMA dataset ${{D}_{IMA}}$ that can perform DP.
\State Set $i:=0$
\While{$\left| D \right|\ge 2k$}
    \State Computing the boundary point $P$ and ${P}'$.
    \State ${{C}_{i}}\leftarrow k$ nearest instances to $P$ from $D$ according to $ Dist\left( x,y \right)$, $D:=D\backslash {{C}_{i}}$.
    \State ${{C}_{i+1}}\leftarrow $k nearest instances to ${P}'$ from $D$ according to $ Dist\left( x,y \right)$, $D:=D\backslash {{C}_{i+1}}$.
    \State $i:=i+2$
\EndWhile
\State ${{C}_{i}}\leftarrow $ remaining records.
\State Computing each centroid of ${{C}_{i}}$ and use it to replace the records in each cluster.\\
\Return ${{D}_{IMA}}$.
\end{algorithmic}
\end{algorithm}

\subsubsection{Determining the sensitivity}
As Definition \ref{def:Sensitivity}, $\Delta {{f}_{j}}=max\left( {{A}_{j}} \right)$ in dataset $D$,
and the sensitivity in IMA is $\frac{\Delta {{\text{f}}_{j}}}{k}\times \frac{m}{2k}$, which is formalized in the Proposition \ref{proposition:the Sensitivity in IMA}.
We detail Algorithm \ref{algorithm:IMA e-DP Mechanism} constructing our DP mechanism.

\begin{myproposition}[$\Delta f$ in IMA]\label{proposition:the Sensitivity in IMA}
${{f}_{j}}\left( D \right)$ is a query function with $\epsilon $-DP mechanism returning the noised values corresponding to the j-th attribute of $D$.
After obtaining ${{D}_{IMA}}$ by Algorithm \ref{algorithm:IMA}, the sensitivity of ${{D}_{IMA}}$ with cluster size $k$ is $\Delta {{{f}'}_{j}}\left( D \right)=\frac{\Delta {{\text{f}}_{j}}}{k}\times \frac{m}{2k}$, where $\Delta {{\text{f}}_{j}}=\text{max}\left( {{A}_{j}} \right)$.	
\end{myproposition}

\begin{myproof}[Proof of Proposition \ref{proposition:the Sensitivity in IMA}]\label{myproof:the Sensitivity in IMA}
If $M$ is an IMA algorithm, for every pair of datasets $D$ and $D'$ differing in a single record, there is a bijection between the set of clusters $\left\{ {{C}_{1}},\ldots ,{{C}_{n}} \right\}$ and $\left\{ C_{1}^{'},...,C_{n}^{'} \right\}$ such that each pair of corresponding clusters differs at most in a single record.
So if the centroid is computed as the mean of the records in the same cluster, then the maximum change in any centroid is, at most, $\frac{\Delta {{f}_{j}}}{k}$.
The modification of single record may lead to multiple modifications of the centroid of clusters, and there are $\left \lceil \frac{m}{k} \right \rceil$\footnote{$\left \lceil \  \right \rceil$ denotes a ceiling functions.} different clusters in $D$.
\\\indent
According to a distance function $Dist()$ with an total order relation,
IMA algorithm iteratively takes sets with cardinality $k$ from the extreme points until less than $2k$ records are left.
The less than $2k$ records are formed the last cluster ${{C}_{r}}$ that is the cluster at the center of the total order relation sequence.
Every ${{x}_{i}}$ in $D$ is ordered by $Dist()$.
A pair of databases $\left( D,D' \right)$ differing only in one instance ${{x}_{i}}$ means the larger database contains just one additional row \cite{18}.
The number of clusters on the left and the right of ${{C}_{r}}$ is equal,
as shown in the Figure \ref{fig:Cluster in IMA}.
If the different record in $\left( D,D' \right)$ is $x_{i}$ on the left of ${{C}_{r}}$ and ${{x}_{i}}$ is located to ${{C}_{i}}$.
Then the changed clusters are the clusters from ${{C}_{i}}$ to ${{C}_{r}}$, and the maximum change for each changed cluster is $\frac{\Delta {{\text{f}}_{j}}}{k}$.
%there will be a maximum $\frac{\Delta {{\text{f}}_{j}}}{k}$ change on each changed cluster centroid from the cluster ${{C}_{i}}$ to ${{C}_{r}}$.
Other clusters on the right side of ${{C}_{r}}$ will not be changed.
The worst scenario is when ${{x}_{i}}$ is located to ${{C}_{1}}$, there is the maximum number $\left \lceil \frac{m}{2k} \right \rceil$ of changed clusters.
The scenario on the left and the right sides of $C_{r}$ is symmetrical,
so the number of changed clusters is at most $\left \lceil \frac{m}{2k} \right \rceil$.
%, then $\Delta {f'}_{j} \left( D \right)\le \frac{\Delta {{f}_{j}}}{k}\times \frac{m}{2k}$.
\begin{figure}[htb]
\centering
\includegraphics[width=8.8cm]{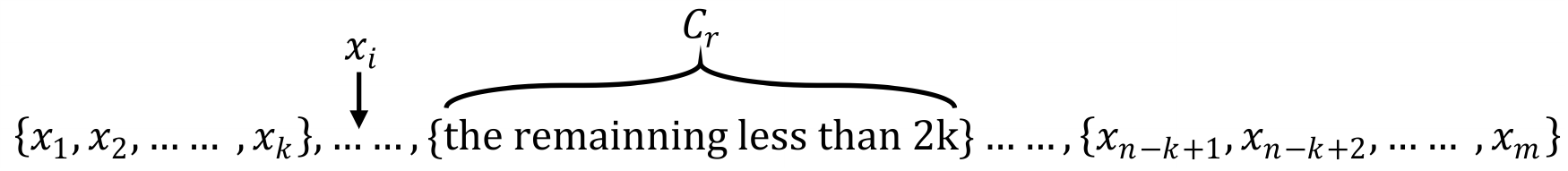}
\caption{Clusters in IMA}\label{fig:Cluster in IMA}
\end{figure}
$\hfill\square$
\end{myproof}

To make the sensitivity of $D_{IMA}$ smaller than the original dataset, we let $\frac{\Delta {{f}_{j}}}{k}\times \frac{m}{2k} \le \Delta {f}_{j}$,
then we can get the best cluster size: $\text{k}=\sqrt[{}]{\frac{m}{2}}$.
Soria et al. \cite{18} thought that $D'$ and $D$ differ in a ``modified" record $x_{i}$.
The modification causes the whole sequence originally obtained by $Dist()$ is changed from the position of $x_{i}$ in turn.
So they considered the sensitive is $\Delta {{{f}'}_{j}}\left( D \right)=\frac{\Delta {{\text{f}}_{j}}}{k}\times \frac{m}{k}$.
However, Dwork er al. \cite{18} give that:
``On pairs of Adjacent Dataset $(D, D')$ differing only in one row, meaning one is a subset of the other and the larger database contains just one additional row."
Their sensitive method causes a greater sensitivity than ours,
which reduces the usability of the dataset.

%that combines Algorithm \ref{algorithm:generating the appropriate value of e} to obtain appropriate $\epsilon$ and Algorithm \ref{algorithm:IMA} to obtain the IMA clusters.
\begin{algorithm}[!t]
\renewcommand\baselinestretch{1.2}\selectfont
%\floatname{algorithm}{Protocol}
\small
\caption{IMA $\epsilon$-DP Mechanism}\label{algorithm:IMA e-DP Mechanism} % 算法的名字
\begin{algorithmic}[1]
\Require \textbf{:} $D=\left\{ \left( {{x}_{i}},{{y}_{i}} \right) \right\}_{i=1}^{m}$ ,k is the cluster size, $m\ge 2k$.
\Ensure \textbf{:} An IMA $\epsilon$-DP dataset ${{D}_{IMA-\epsilon }}$.
\State Generating the appropriate $\epsilon $ on dataset $D$ by Algorithm \ref{algorithm:generating the appropriate value of e}.
\State Obtaining an IMA dataset ${{D}_{IMA}}$ from $D$ by Algorithm \ref{algorithm:IMA}.
\State Obtaining noise by using $\epsilon $ and $\Delta {{f}_{j}}$ (cf. the definition of Laplace mechanism \ref{def:Laplace mechanism}).
\State Adding $x_{i}^{\epsilon }={{x}_{i}}+noise$ to ${{D}_{IMA-\epsilon }}$.
\\
\Return ${{D}_{IMA-\epsilon }}$.
\end{algorithmic}
\end{algorithm}

\section{Privacy Design with Homomorphic Cryptosystem}\label{sec:HC}
A homomorphic encryption algorithm can only support one type of operation (e.g., Addition or Multiplication).
Existing HC-based secure training algorithms need to rely on trusted third parties such as the Authorization Server \cite{8,20}, or use an approximate equation to simplify the original complex iteration formula\cite{36,46}.

We elaborately design a library of building blocks by multiplicative homomorphic encryption RSA and additively homomorphic encryption Paillier,
which is able to construct complex secure ML training algorithms needing no the Authorization Server or any approximate equation.
In order to illustrate the power of our building blocks, we construct a secure LR training algorithm (see Section \ref{sec:combine-Constructing a Specific Training Algorithms using Building Blocks}).
It is the first solving the sigmoid function based on HC.
In this section, we detail our building blocks.
The security proofs for each building block are given in Section \ref{sec:Security Analysis}.

ML training algorithms are computationally complex,
so the building blocks need to support a range of choices including which party gets the input, which party gets the output, and whether the input or output are encrypted or not.
Table \ref{table:The Conditions of Building Blocks} shows the different conditions for building blocks.
For all conditions, both parties Alice and Bob cannot obtain other useful information except for the legal information,
the input and output of other parties are confidential.
%We describe the building blocks we design in detail below followed with their corresponding conditions.
%Our building blocks (except Protocol \ref{protocol:Secure dot product 2}) shown below are the operation of vectors,
%if an integer operation is required, it can be regarded as a special case where the vector is one-dimensional.

\begin{table*}[!t]
\normalsize
\centering
\small
\caption{The Conditions of Building Blocks}\label{table:The Conditions of Building Blocks}
\renewcommand{\arraystretch}{1.1}
\begin{tabular}{|c|c|c|c|c|c|}
\hline
\multirow{2}{*}{Conditions}&\multicolumn{2}{c|}{Input}&\multicolumn{2}{c|}{Output}&\multirow{2}{*}{Protocols}	
\\
\cline{2-5}	
 & \multicolumn{1}{c}{Alice}&	 \multicolumn{1}{c|}{Bob}&	\multicolumn{1}{c}{ Alice}&\multicolumn{1}{c|}{  Bob}&\multicolumn{1}{c|}{}
\\
\hline
\hline
Condition\ 1&	$\sf{SK}_{A}$, $\sf{PK}_{B}$, $a$  &	$\sf{SK}_{B}$, $\sf{PK}_{A}$, $b$	  &  -&${[ f(a,b) ]}_{A}$&
\ref{protocol:Secure Addition Protocol}, \ref{protocol:Secure Subtraction Protocol}, \ref{protocol:Secure dot product}, \ref{protocol:Security exponential Power}
\\\hline
Condition\ 2	& $\sf{SK}_{A}$, $\sf{PK}_{B}$, $a$	&  $\sf{SK}_{B}$, $\sf{PK}_{A}$, ${{\left[ b \right]}_{A}}$	&  -&${[ f(a,b) ]}_{A}$&
\ref{protocol:Secure Addition Protocol}, \ref{protocol:Secure Subtraction Protocol}, \ref{protocol:Secure dot product 2}
\\\hline
Condition\ 3	& $\sf{SK}_{A}$, $\sf{PK}_{B}$ &	$\sf{SK}_{B}$, $\sf{PK}_{A}$, ${{|| f(a,b) ||}_{A}}$& - & ${{[[f(a,b)]]}}_{A}$ &
\ref{protocol:Converting Ciphertext 1}
\\\hline
Condition\ 4	&$ \sf{SK}_{A}$, $\sf{PK}_{B}$&	$\sf{SK}_{B}$, $\sf{PK}_{A}$, ${[[f(a,b)]]}_{A}$	&	-& ${[[ f(a,b) ]]}_{B}$&
\ref{protocol:Converting Ciphertext 2}
\\
\hline
\end{tabular}
\end{table*}

\textbf{\textit{1) Secure addition and secure subtraction:}}
Relying on Paillier's additive homomorphic property, it is straightforward to obtain the secure addition protocol (Protocol \ref{protocol:Secure Addition Protocol}) and secure subtraction protocol (Protocol \ref{protocol:Secure Subtraction Protocol}).
%that satisfy the Conditions 1 and 4 in the Table \ref{table:The Conditions of Building Blocks} under Paillier.

%\begin{myproposition}\label{proposition:Secure addition}
%Protocol \ref{protocol:Secure Addition Protocol} is secure in the honest-but-curious model.
%\end{myproposition}

%\begin{myproposition}\label{proposition:Secure subtraction}
%Protocol \ref{protocol:Secure Subtraction Protocol} is secure in the honest-but-curious model.
%\end{myproposition}

\begin{algorithm}[!t]
\renewcommand\baselinestretch{1.2}\selectfont
\floatname{algorithm}{Protocol}
\setcounter{algorithm}{0}
\small
\caption{Secure Addition Protocol}\label{protocol:Secure Addition Protocol} % 算法的名字
\begin{algorithmic}[1]
\Require \textbf{Alice:} $a=\left\{ {{a}_{1}},{{a}_{2}},\ldots \ldots,{{a}_{d}} \right\}$, ($\sf{PK}_{Alice}$, $\sf{SK}_{Alice}$).
\Require \textbf{Bob:} $\sf{PK}_{Alice}$, ${{[[b]]}_{Alice}}={{\left[\left[ \left\{ {{b}_{1}},{{b}_{2}},\ldots \ldots,{{b}_{d}} \right\} \right]\right]}_{Alice}}$ or  b=$\left\{ {{b}_{1}},{{b}_{2}},\ldots \ldots,{{b}_{d}} \right\}$.
\Ensure \textbf{Bob:} ${{\left[\left[ f\left( a,b \right) \right]\right]}_{Alice}}={{\left[\left[ a+b \right]\right]}_{Alice}}$.
\State Alice sends ${{\left[\left[ a \right]\right]}_{Alice}}$ to Bob.
\For{i to d}
    \State Bob computes ${{\left[\left[ f{{\left( a,b \right)}_{i}} \right]\right]}_{Alice}}={{\left[\left[ {{b}_{i}} \right]\right]}_{Alice}}\times {{\left[\left[ {{a}_{i}} \right]\right]}_{Alice}}$.  \label{protocol:Secure Addition Protocol-important}
\EndFor\\
\Return ${{\left[\left[ f\left( a,b \right) \right]\right]}_{Alice}}$ to Bob.
\end{algorithmic}
\end{algorithm}

\begin{algorithm}[!t]
\renewcommand\baselinestretch{1.2}\selectfont
\floatname{algorithm}{Protocol}
\small
\caption{Secure Subtraction Protocol}\label{protocol:Secure Subtraction Protocol} % 算法的名字
\begin{algorithmic}[1]
\Require \textbf{Alice:} $a=\left\{ {{a}_{1}},{{a}_{2}},\ldots \ldots,{{a}_{d}} \right\}$, ($\sf{PK}_{Alice}$, $\sf{SK}_{Alice}$).
\Require \textbf{Bob:} $\sf{PK}_{Alice}$, ${{[[b^{-1}]]}_{Alice}}={{[[ \left\{ {{b}_{1}^{-1}},{{b}_{2}^{-1}},\ldots ,
{{b}_{d}^{-1}} \right\} ]]}_{Alice}}$ or $b^{-1}=\left\{ {{b}_{1}^{-1}},{{b}_{2}^{-1}}, \ldots,{{b}_{d}^{-1}} \right\}$.
\Ensure \textbf{Bob:} ${{\left[\left[ f\left( a,b \right) \right]\right]}_{Alice}}={{\left[\left[ a-b \right]\right]}_{Alice}}$.
\State Alice sends ${{\left[\left[ a \right]\right]}_{Alice}}$ to Bob.
\For{i to d}
    \State Bob computes ${{\left[\left[ f{{\left( a,b \right)}_{i}} \right]\right]}_{Alice}}={{\left[\left[ {{b}_{i}^{-1}} \right]\right]}_{Alice}}\times {{\left[\left[ {{a}_{i}} \right]\right]}_{Alice}}$.
\EndFor\\
\Return ${{\left[\left[ f\left( a,b \right) \right]\right]}_{Alice}}$ to Bob.
\end{algorithmic}
\end{algorithm}

\textbf{\textit{2) Secure dot product and secure multiplication:}}
Using Paillier's additive homomorphism property, we can construct a secure dot product protocol (Protocol \ref{protocol:Secure dot product}) that satisfies Condition 1 easy (i.e., $\left[\left[ a\times b \right]\right]={{\left[\left[ a \right]\right]}^{b}}$).
However, when Bob only has ciphertext ${{\left[\left[ b \right]\right]}_{Alice}}$ who is unable to perform $\left[\left[ a \right]\right]_{Alice}^{b}$.
Paillier fails to construct a secure dot product protocol that satisfies Condition 2.
%One would say $\left[\left[ a\times b \right]\right]={{\left[\left[ a \right]\right]}^{\left[\left[ b \right]\right]}}$,
Because the length of Paillier's ciphertext is $1024$ bits or longer usually (We will discuss the key length setting in detail in Section \ref{sec:PE}.),
so the computational complexity of ${{\left[\left[ a \right]\right]}^{\left[\left[ b \right]\right]}}$ is awfully large.
%and it is unable calculated for a limited time.
Therefor, when faced with Condition 2, we use RSA's multiplicative homomorphism property to construct secure multiplication protocol (Protocol \ref{protocol:Secure dot product 2}).

%\begin{myproposition}\label{proposition:Secure dot product 1}
%Protocol \ref{protocol:Secure dot product 1} is secure in the honest-but-curious model.
%\end{myproposition}

%\begin{myproposition}\label{proposition:Secure dot product 2}
%Protocol \ref{protocol:Secure dot product 2} is secure in the honest-but-curious model.
%\end{myproposition}

\begin{algorithm}[!t]
\renewcommand\baselinestretch{1.2}\selectfont
\floatname{algorithm}{Protocol}
\small
\caption{Secure Dot Product Protocol}\label{protocol:Secure dot product} % 算法的名字
\begin{algorithmic}[1]
\Require \textbf{Alice:} $a=\left\{ {{a}_{1}},{{a}_{2}},\ldots \ldots,{{a}_{d}} \right\}$, ($\sf{PK}_{Alice}$, $\sf{SK}_{Alice}$).
\Require \textbf{Bob:} $b=\left\{ {{b}_{1}},{{b}_{2}},\ldots \ldots,{{b}_{d}} \right\}$ and $\sf{PK}_{Alice}$.
\Ensure \textbf{Bob:} ${{\left[\left[ f\left( a,b \right) \right]\right]}_{Alice}}={{\left[\left[ a\times b \right]\right]}_{Alice}}$.
\State Alice sends ${{\left[\left[ a \right]\right]}_{Alice}}$ to Bob.
\State Bob computes ${{\left[\left[ f\left( a,b \right) \right]\right]}_{Alice}}=\underset{i=1}{\overset{d}{\mathop \prod }}\,{{\left[\left[ {{a}_{i}} \right]\right]}_{Alice}^{{b}_{i}}}$.\\
\Return ${{[[ f ( a,b) ]]}_{Alice}}$ to Bob.
\end{algorithmic}
\end{algorithm}

\begin{algorithm}[!t]
\renewcommand\baselinestretch{1.2}\selectfont
\floatname{algorithm}{Protocol}
\small
\caption{Secure Multiplication Protocol}\label{protocol:Secure dot product 2} % 算法的名字
\begin{algorithmic}[1]
\Require \textbf{Alice:} $a$, ($\sf{PK}_{Alice}$, $\sf{SK}_{Alice}$).
\Require \textbf{Bob:} ${|| b ||}_{Alice}$ and $\sf{PK}_{Alice}$.
\Ensure \textbf{Bob:} ${{|| f( a,b ) ||}_{Alice}}={{|| a\times b ||}_{Alice}}$.
\State Alice sends ${{|| a ||}_{Alice}}$ to Bob.
\State Bob computes ${{||f(a,b) ||}_{Alice}}= {||a||}_{Alice}\times {||b||}_{Alice}$.\\
\Return ${{|| f ( a,b) ||}_{Alice}}$ to Bob.
\end{algorithmic}
\end{algorithm}

\textbf{\textit{3) Secure power function:}}
In order to cope with more complex training algorithms, we design protocol \ref{protocol:Security exponential Power} satisfying Condition 1 under RSA to obtain ${||{e}^{ab}||}$ securely.
 %where $a=\left\{ {{a}_{1}},{{a}_{2}},\ldots \ldots,{{a}_{d}} \right\}$ and $b=\left\{ {{b}_{1}},{{b}_{2}},\ldots \ldots,{{b}_{d}} \right\}$.

%\begin{myproposition}\label{proposition:Security exponential Power}
%Protocol \ref{protocol:Security exponential Power} is secure in the honest-but-curious model.
%\end{myproposition}

\begin{algorithm}[!t]
\renewcommand\baselinestretch{1.2}\selectfont
\floatname{algorithm}{Protocol}
\small
\caption{Secure Power Function Protocol}\label{protocol:Security exponential Power} % 算法的名字
\begin{algorithmic}[1]
\Require \textbf{Alice:} $a=\left\{ {{a}_{1}},{{a}_{2}},\ldots \ldots,{{a}_{d}} \right\}$, ($\sf{PK}_{Alice}$, $\sf{SK}_{Alice}$).
\Require \textbf{Bob:} $b=\left\{ {{b}_{1}},{{b}_{2}},\ldots \ldots,{{b}_{d}} \right\}$ and $\sf{PK}_{Alice}$.
\Ensure \textbf{Bob:} ${{\left\| {{e}^{ab}} \right\|}_{Alice}}$.
\State Alice sends ${{\left\| {{e}^{a}} \right\|}_{Alice}}={{\left\| \left\{ {{e}^{{{a}_{1}}}},{{e}^{{{a}_{2}}}},\ldots {{e}^{{{a}_{d}}}} \right\} \right\|}_{Alice}}$ to Bob.
\State Bob Initializes $W$.
\State In Bob:
\For{i to d}
    \State Letting ${{w}_{i}}={{\left\| {{e}^{{{x}_{i}}}} \right\|}_{Alice}}$.
	\For{t to ${{b}_{i}}-1$}
       \State ${{w}_{i}}={{w}_{i}}\times {{\left\| {{e}^{{{a}_{i}}}} \right\|}_{Alice}}$ by protocol 4.
    \EndFor
    \State $W=W\times {{w}_{i}}$ by protocol 4.
\EndFor
\\
\Return ${{\left\| W \right\|}_{Alice}}={{\left\| {{e}^{ab}} \right\|}_{Alice}}$ to Bob.
\end{algorithmic}
\end{algorithm}

\textbf{\textit{4) Secure changing the encryption cryptosystem:}}
There are multiple participants in \texttt{Heda}.
Different participants have their own encryption scheme (i.e., the certain plaintext space and a pair of keys ($\sf{PK}$,$\sf{SK}$)).
Homomorphic operations can only be operated in the same plaintext space.
For completeness, we design two protocols converting ciphertext from one encryption scheme to another while maintaining the underlying plaintext.
The first (Protocol \ref{protocol:Converting Ciphertext 1}) switches ${{\left\| {{e}^{ab}} \right\|}_{Alice}}$ to ${{\left[\left[ {{e}^{ab}} \right]\right]}_{Alice}}$ satisfying Condition 3,
and the other (Protocol \ref{protocol:Converting Ciphertext 2}) switches ${{\left[\left[ b \right]\right]}_{Alice}}$ to ${{\left[\left[ b \right]\right]}_{Bob}}$ satisfying Condition 4.

%\begin{myproposition}\label{proposition:Secure changing the encryption cryptosystem 1}
%Protocol \ref{protocol:Converting Ciphertext 1} is secure in the honest-but-curious model.
%\end{myproposition}

\begin{myproposition}[The security of building blocks]\label{proposition:The security of building blocks}
Protocol \ref{protocol:Secure Addition Protocol} to \ref{protocol:Converting Ciphertext 2} is secure in the honest-but-curious model.
\end{myproposition}

\begin{algorithm}[!t]
\renewcommand\baselinestretch{1.2}\selectfont
\floatname{algorithm}{Protocol}
\small
\caption{Converting Ciphertext: ${{\left\| {{e}^{ab}} \right\|}_{Alice}}$ to ${{\left[\left[ {{e}^{ab}} \right]\right]}_{Alice}}$}\label{protocol:Converting Ciphertext 1} % 算法的名字
\begin{algorithmic}[1]
\Require \textbf{Alice:} ($\sf{PK}_{Alice-Paillier}$, $\sf{SK}_{Alice-Paillier}$) and ($\sf{PK}_{Alice-RSA}$, $\sf{SK}_{Alice-RSA}$).
\Require \textbf{Bob:} ${{\left\| {{e}^{ab}} \right\|}_{Alice}}$.
\Ensure \textbf{Bob:} ${{\left[\left[ {{e}^{ab}} \right]\right]}_{Alice}}$.
\State Bob uniformly picks $r$ and computes ${{\left\| {{e}^{ab+r}} \right\|}_{Alice}}$ by Protocol 5.
\State Bob sends ${{\left\| {{e}^{ab+r}} \right\|}_{Alice}}$ to Alice.
\State Alice decrypts ${{\left\| {{e}^{ab+r}} \right\|}_{Alice}}$, obtains and sends ${{\left[\left[ {{e}^{ab+r}} \right]\right]}_{Alice}}$ to Bob.
\State Bob computes ${{\left[\left[ {{e}^{ab}} \right]\right]}_{Alice}}={{\left[\left[ {{e}^{ab+r}} \right]\right]}_{Alice}}\times {(e^{r})^{-1}}$ by protocol 3.\\
\Return ${{\left[\left[ {{e}^{ab}} \right]\right]}_{Alice}}$ to Bob.
\end{algorithmic}
\end{algorithm}

\begin{algorithm}[!t]
\renewcommand\baselinestretch{1.2}\selectfont
\floatname{algorithm}{Protocol}
\small
\caption{Converting Ciphertext: ${{\left[\left[ b \right]\right]}_{Alice}}$ to ${{\left[\left[ b \right]\right]}_{Bob}}$}\label{protocol:Converting Ciphertext 2} % 算法的名字
\begin{algorithmic}[1]
\Require \textbf{Alice:} ($\sf{PK}_{Alice}$, $\sf{SK}_{Alice}$) and $\sf{PK}_{Bob}$.
\Require \textbf{Bob:} ${{\left[\left[ b \right]\right]}_{Alice}}$.
\Ensure \textbf{Bob:} ${{\left[\left[ b \right]\right]}_{Bob}}$.
\State Bob uniformly picks $r$ and computes ${{\left[\left[ b+r \right]\right]}_{Alice}}$ by Protocol 1.
\State Bob sends ${{\left[\left[ b+r \right]\right]}_{Alice}}$ to Alice.
\State Alice decrypts ${{\left[\left[ b+r \right]\right]}_{Alice}}$, obtains and sends ${{\left[\left[ b+r \right]\right]}_{Bob}}$ to Bob.\\
\Return ${{\left[\left[ b \right]\right]}_{Bob}}$ to Bob.
\end{algorithmic}
\end{algorithm}

\section{Construction of \texttt{Heda}}\label{sec:combine}
%In Section \ref{sec:PS-System model}, we indicate \texttt{Heda}'s idea in three steps: feature evaluation; obtaining model parameters by building blocks and DP; obtaining overall model parameters.
%These three parts is designed in a modular way, carrying out secure ML training algorithm come down to invoking the right module.
%The part of DP and HC in \texttt{Heha} have been described in Section \ref{sec:DP} and \ref{sec:HC}, respectively.
In this section, we introduce the overall framework of \texttt{Heda}.
In \texttt{Heda}, $\mathcal{U}$ is able to learn a model without learning anything about the sensitive data of $\mathcal{P}$,
and in addition to $\mathcal{U}$, others should learn nothing about the model.
The security proof will be given in Section \ref{sec:Security Analysis}.

\texttt{Heda} is exhibited in Algorithm \ref{algorithm:Privacy-Preserving Training}.
We introduce the following details in this section:
how to use feature evaluation technologies to divide a dataset into two parts (Algorithm \ref{algorithm:Privacy-Preserving Training} Step 2),
how to construct a specific training algorithm using building blocks (Algorithm \ref{algorithm:Privacy-Preserving Training} Step 4),
and how to combine DP mechanism with the building blocks (Algorithm \ref{algorithm:Privacy-Preserving Training} Step 4).

\begin{algorithm}[!t]
\setcounter{algorithm}{3}
\renewcommand\baselinestretch{1.2}\selectfont
%\floatname{algorithm}{Protocol}
\small
\caption{Privacy-Preserving Training}\label{algorithm:Privacy-Preserving Training} % 算法的名字
\begin{algorithmic}[1]
\renewcommand{\algorithmicrequire}{\textbf{Each $\mathcal{P}$'s Input:}}
\Require  ${{D}_{i}}=\left\{ \left( {{x}_{i}},{{y}_{i}} \right) \right\}_{i=1}^{m}$, ($\sf{PK}_{{{\mathcal{P}}_{i}}-{Paillier}}$,$\sf{SK}_{{{\mathcal{P}}_{i}}-Paillier}$) and ($\sf{PK}_{{{\mathcal{P}}_{i}}-{RSA}}$,$\sf{SK}_{{{\mathcal{P}}_{i}}-RSA}$).
\renewcommand{\algorithmicrequire}{\textbf{$\mathcal{U}$'s Input:}}
\Require ($\sf{PK}_{\mathcal{U}-Paillier}$, $\sf{SK}_{\mathcal{U}-Paillier}$).
\renewcommand{\algorithmicensure}{\textbf{$\mathcal{U}$'s Output:}}
\Ensure $\beta$.
\State $\mathcal{U}$ initializes $\beta$.
%\State $\mathcal{U}$ obtains $S$ by Algorithm \ref{algorithm:Secure Scoring Features}.
\State According to $S$, each $\mathcal{P}$ divides all features in two part: the high scores part and the low scores part.
\State each $\mathcal{P}$ obtains the noise dataset from the low scores part subdataset by Algorithm \ref{algorithm:IMA e-DP Mechanism} and sends it to $\mathcal{U}$.
\State $\mathcal{U}$ trains a model by building blocks (Algorithm \ref{algorithm:Training Private-Preserving LR}) combine with noised datasets.\\
\Return $\beta$ to $\mathcal{U}$.
\end{algorithmic}
\end{algorithm}

\begin{myproposition}[The Security of \texttt{Heda}]\label{proposition:the security of heda}
Algorithm \ref{algorithm:Privacy-Preserving Training} is secure in the honest-but-curious model.
\end{myproposition}

%\begin{figure}[htb]
%\centering
%\includegraphics[width=6.5cm]{fig/SystemModel}
%\caption{The system model of \texttt{Heda}.}\label{fig:SystemModel}
%\end{figure}

\subsection{Feature Partitioning}\label{sec:combine-Dividing a Aataset into Two Parts According to the Feature Scores}
%In this section, we describe how to implement secure feature scoring when facing multi-party encrypted data.
%The security analysis gives in Section \ref{sec:Security Analysis}.
Obviously, It is best that each $\mathcal{P}$ conducts the feature evaluation locally.
The locally computation does not require interaction with any other party which guarantee the privacy of each $\mathcal{P}$.
In addition, $\mathcal{P}$ can perform arbitrary computations on their sensitive data in plaintext case with high efficiency.
%Several $\mathcal{P}$ involve in \texttt{Heda}.
After Several $\mathcal{P}$ who join in \texttt{Heda} locally implement feature evaluation,
they communicate with each other negotiating the final scores $S$.
Feature scores do not reveal the privacy of datasets, so it is feasible that several $\mathcal{P}$ share the scores of their datasets and negotiate the final feature scores.

According to the feature scores $S$, each $\mathcal{P}$ processes the dataset into an ordered dataset.
Let an ordered dataset $\overset{}{\mathop{D}}\,=\left\{ {{A}_{1}},{{A}_{2}},..{{A}_{d}} \right\}$ ordered by feature scores, i.e. ${{S}_{i}}$ is the scores of ${{A}_{i}}$, ${{S}_{i}}>{{S}_{j}},0<i<j<d$.
Let the high scores part has $ \iota $ features, then $\overset{}{\mathop{D}}\,=\left\{ {{\overset{}{\mathop{D}}\,}_{1}},{{\overset{}{\mathop{D}}\,}_{2}} \right\}$, ${{\overset{}{\mathop{D}}\,}_{1}}=\left\{ {{A}_{1}},{{A}_{2}},..{{A}_{ \iota }} \right\}$, ${{\overset{}{\mathop{D}}\,}_{2}}=\left\{ {{A}_{ \iota +1}},{{A}_{ \iota +2}},..{{A}_{d}} \right\}$.

We assume that $\mathcal{U}$ spends ${{t}_{1}}$ in learning a classifier parameters on the low scores part (the noise dataset)
and ${{t}_{2}}$ on the high scores part (the encrypted dataset),
then the total time is $T={{t}_{1}}+{{t}_{2}}$.
Training a model in plaintext case usually takes time in milliseconds
but usually takes thousands of seconds or longer in ciphertext case \cite{20,9}.
There is a linear relationship between ${{t}_{2}}$ and $\iota$, i.e. ${{t}_{2}}\approx \tau \iota + b$, where $ \iota >0$,
$\tau $ and $b$ are two constants.
The training time on noise dataset is much less than the time training a model on the encrypted dataset.
Formula \ref{equ:total time consumption} shows the total time consumption.
\begin{equation}\label{equ:total time consumption}
  T\le \tau \left(  \iota +1 \right) +b
\end{equation}
\indent
\texttt{Heda} enables flexible switch among different tradeoffs between efficiency and accuracy by parameter tuning.
With parameter $ \iota $, one is able to obtain the tradeoff between efficiency and accuracy.
As the decreasing number of the $ \iota $, the total time consumption is consequent reduction.
When the number of dimensions assigned to the high scores part is small, the accuracy is relatively low.
According to the specific situations, $\iota $ is set appropriately.

As for the selection of feature evaluation techniques,
many excellent feature evaluation techniques have been studied \cite{6,48}.
When facing a dataset with different types, different backgrounds or different magnitude, different methods have their drawbacks as well as merits.
We evaluate six widely used methods in our experiments.
The methods we use are:
Chi-square test,
Kruskal-Wallis H (KW),
Pearson correlation,
Spearman correlation,
Random forest and
minimal Redundancy Maximal Relevance (mRMR).
We are committed to finding a feature evaluation technique with the best robustness.
After extensive experiments, we find that KW has the most stable effect when facing with different datasets (see Section \ref{sec:PE-Performance of Combining HC and DP}).

\subsection{Constructing Specific Training Algorithms using Building Blocks}\label{sec:combine-Constructing a Specific Training Algorithms using Building Blocks}
There are rich variety of ML algorithms.
Describing the implementation of building blocks towards each ML training algorithm naturally requires space beyond the page limit.
We use LR\footnote{In order to maintain the continuity of our description, LR is also used as the example in the following} as an example to illustrate how to construct secure model training algorithms by our building blocks.
%It is worth mentioning that our goal is not to construct just a LR model but various ML training algorithms,
%and our building blocks can also build other training algorithms.

LR training algorithm is not the most complicated one compared to other ML classifier training algorithms.
However, the iterative process of LR involves sigmoid function ($\text{Sigmoid}\left( {{\beta }^{T}}{{x}_{i}} \right)=\frac{{{e}^{{{\beta }^{T}}{{x}_{i}}}}}{1+{{e}^{{{\beta }^{T}}{{x}_{i}}}}}$) which makes it difficult to implement in ciphertext case.
Most studies claimed they had constructed a secure LR training algorithm by HC which were the secure linear regression training algorithms actually,
or they solved the sigmoid function by an approximate equation (cf. Section \ref{sec:RW-Privacy-Preserving ML Classifier Training}).
To best of our knowledge, it is the first constructing a secure LR training algorithm by HC.
Our HC-based secure LR training algorithm only needs 3 interactions (i.e. interactions between $\mathcal{U}$ and $\mathcal{P}$) throughout each iteration process.

LR is a binary classifier and try to learn a pair of parameters $w$ and $b$,
where $w=\left( {{w}_{1}},{{w}_{2}},\ldots,{{w}_{d}} \right)$ to satisfy $f\left\{ {{x}_{i}} \right\}={{w}^{T}}{{x}_{i}}+b$ and $f \left\{ {{x}_{i}} \right\}\cong {{y}_{i}}$.
LR uses Sigmoid Function to associate the true label ${{y}_{i}}$ with the prediction label $f \left\{ {{x}_{i}} \right\}$.
%The $y$ in $\text{Sigmoid}\left( y \right)$ is usually regarded as a posterior probability estimate$~P \left( Y=1|x \right)$: $ln \frac{P\left( y=1|x \right)}{P\left( y=1|x \right)}={{w}^{T}}{{x}_{i}}+b$, and Maximum Likelihood Method with gradient descent method is commonly used to estimate the value of $w$ and $b$ and to find optimal solution.
Let $\beta =(w,b)$, $X=\left( x,1 \right)$, the iteration formula of LR is shown in Formula \eqref{equ:iteration formula of LR}.
The steps of LR training algorithm are as follows:
(i) Initializing learning rate $\alpha $, a fixed number of iterations and model parameters $\beta$.
(ii) Updating $\beta$ by Formula \eqref{equ:iteration formula of LR}.
(iii) If the maximum number of iterations or minimum learning rate is reached, output $\beta$; otherwise go to step 2.
%For more details about LR, we refer the reader to \cite{57}.
%Formula \eqref{equ:iteration formula of LR} contains operations for subtraction and division whose results exceed the cipher space of Paillier or RSA that cannot be easily obtained through homomorphic properties.
\begin{equation}\label{equ:iteration formula of LR}
  \beta _{j}^{t+1}=~\beta _{j}^{t}-\frac{\alpha }{m}\underset{i=1}{\overset{m}{\mathop \sum }}\,{{x}_{ij}}\left( \frac{{{e}^{{{\beta }^{T}}{{x}_{i}}}}}{1+{{e}^{{{\beta }^{T}}{{x}_{i}}}}}-{{y}_{i}} \right)
\end{equation}

Each building block is designed in a modular way, so carrying out secure LR training algorithm come down to invoking the right module.
Suppose there are n data providers $\mathcal{P}$, Algorithm \ref{algorithm:Training Private-Preserving LR} specifies our secure LR training algorithm.
In all execution steps of Algorithm \ref{algorithm:Training Private-Preserving LR},
when protocols are called,
$\mathcal{P}$ is the role of Alice, and $\mathcal{U}$ is the role of Bob.
%In each iteration, the $\mathcal{U}$ only needs to interact with each $\mathcal{P}$ three times.

\begin{myproposition}\label{proposition:Training Private-Preserving LR}
Algorithm \ref{algorithm:Training Private-Preserving LR} is secure in the honest-but-curious model.
\end{myproposition}

\begin{algorithm}[!t]
\renewcommand\baselinestretch{1.2}\selectfont
%\floatname{algorithm}{Protocol}
\small
\caption{privacy-preserving training of LR}\label{algorithm:Training Private-Preserving LR} % 算法的名字
\begin{algorithmic}[1]
\renewcommand{\algorithmicrequire}{\textbf{Each $\mathcal{P}$'s Input:}}
\Require  ${{D}_{i}}=\left\{ \left( {{x}_{i}},{{y}_{i}} \right) \right\}_{i=1}^{m}$, ($\sf{PK}_{{{\mathcal{P}}_{i}}-{Paillier}}$,$\sf{SK}_{{{\mathcal{P}}_{i}}-Paillier}$) and ($\sf{PK}_{{{\mathcal{P}}_{i}}-{RSA}}$,$\sf{SK}_{{{\mathcal{P}}_{i}}-RSA}$).
\renewcommand{\algorithmicrequire}{\textbf{$\mathcal{U}$'s Input:}}
\Require ($\sf{PK}_{\mathcal{U}-Paillier}$, $\sf{SK}_{\mathcal{U}-Paillier}$).
\renewcommand{\algorithmicensure}{\textbf{$\mathcal{U}$'s Output:}}
\Ensure $\beta$.
\State $\mathcal{U}$ initializes a learning rate $\alpha $, a fixed number of iterations $Cycles$ and $\beta$.
\While{t in $Cycles$ or minimum learning rate is not reached}
    \State $\mathcal{P}_{i}$ sends $({{\left[\left[ \underset{i=1}{\overset{m}{\mathop \sum }}\,XY \right]\right]}_{{{\mathcal{P}}_{i}}}}$) to $\mathcal{U}$.
    \For{$i=1$ to n}
         \State $\mathcal{P}_{i}$ sends $({{\left\| {{e}^{X}} \right\|}_{{{\mathcal{P}}_{i}}}},{{\left[\left[ \underset{i=1}{\overset{m}{\mathop \sum }}\,XY \right]\right]}_{{{\mathcal{P}}_{i}}}}$) to $\mathcal{U}$.
        %\rightline{$\Rightarrow$$\mathcal{P}_{i}\to \mathcal{U}$.}
        %\State $\mathcal{U}$ computes ${{\left\| {{e}^{\beta X}} \right\|}_{DP{{s}_{i}}}}$ by Protocol \ref{protocol:Security exponential Power}.
        \State $\mathcal{U}$ obtains ${{[[ {e}^{\beta X} ]]}_{\mathcal{P}_{i}}}$ by Protocol \ref{protocol:Security exponential Power} and Protocol \ref{protocol:Converting Ciphertext 1} sequentially.
        %\rightline{$\Rightarrow$ ${$\mathcal{U}$ \to \mathcal{P}_{i}}$, ${\mathcal{P}_{i} \to \mathcal{U}}$.}
        %\State $\mathcal{U}$ obtains ${{\left[\left[ {{e}^{\beta X}}+1 \right]\right]}_{DP{{s}_{i}}}}$ by Protocol \ref{protocol:Secure Addition Protocol}.
        \State $\mathcal{U}$ uniformly picks $r$ and computes ${{\left[\left[ {{e}^{\beta X+r}}+{{e}^{r}} \right]\right]}_{\mathcal{P}_{i}}}$ by Protocol \ref{protocol:Secure Addition Protocol} and Protocol \ref{protocol:Secure dot product 2} sequentially.
        \State $\mathcal{U}$ sends ${{\left[\left[ {{e}^{\beta X+r}}+{{e}^{r}} \right]\right]}_{{{\mathcal{P}}_{i}}}}$ to $\mathcal{P}$.
        %\rightline{$\Rightarrow$$\mathcal{U}\to \mathcal{P}_{i}$.}
        \State $\mathcal{P}$ decrypts ${{\left[\left[ {{e}^{\beta X+r}}+{{e}^{r}} \right]\right]}_{{{\mathcal{P}}_{i}}}}$ and sends ${{\left[\left[ \frac{X}{{{e}^{\beta X+r}}+{{e}^{r}}} \right]\right]}_{\mathcal{P}_{i}}}$ to $\mathcal{U}$.
        %\rightline{$\Rightarrow$$\mathcal{P}_{i}\to \mathcal{U}$.}
        %\State $\mathcal{U}$ obtains ${{\left[\left[ \frac{X}{{{e}^{-\beta X}}+1} \right]\right]}_{\mathcal{P}_{i}}={{\left[\left[ \frac{X}{{{e}^{\beta X+r}}+{{e}^{r}}} \right]\right]}_{\mathcal{P}_{i}}}\times {{e}^{r}}}$ by Protocol \ref{protocol:Secure dot product}.
        \State $\mathcal{U}$ obtains ${{\left[\left[ \frac{X}{{{e}^{-\beta X}}+1} \right]\right]}_{\mathcal{P}_{i}}}$ by Protocol \ref{protocol:Secure dot product}.
        %\State $\mathcal{U}$ obtains ${{\left[\left[ \underset{i=1}{\overset{m}{\mathop \sum }}\,X\frac{{{e}^{\beta X}}}{1+{{e}^{\beta X}}} \right]\right]}_{\mathcal{P}_{i}}}$ by Protocol \ref{protocol:Secure Addition Protocol}.
        \State $\mathcal{U}$ obtains ${{\left[\left[ \underset{i=1}{\overset{m}{\mathop \sum }}\,\left( X\frac{{{e}^{\beta X}}}{1+{{e}^{\beta X}}}-XY \right) \right]\right]}_{\mathcal{P}_{i}}}$ by Protocol \ref{protocol:Secure Addition Protocol} and Protocol \ref{protocol:Secure Subtraction Protocol} sequentially.
        \State $\mathcal{U}$ obtains ${{\left[\left[ \underset{i=1}{\overset{m}{\mathop \sum }}\,\left( X\frac{{{e}^{\beta X}}}{1+{{e}^{\beta X}}}-XY \right) \right]\right]}_{\mathcal{U}}}$ by Protocol \ref{protocol:Converting Ciphertext 2}.
        %\rightline{$\Rightarrow$$\mathcal{U}\to \mathcal{P}_{i}$, $\mathcal{P}_{i}\to \mathcal{U}$.}
        \State $\mathcal{U}$ updates $\beta$ by $\underset{i=1}{\overset{m}{\mathop \sum }}\,\left( X\frac{{{e}^{\beta X}}}{1+{{e}^{\beta X}}}-XY \right)$.
    \EndFor
\EndWhile
\\
\Return $\beta$ to $\mathcal{U}$.
\end{algorithmic}
\end{algorithm}

\subsection{Combining Differential Privacy Mechanism with Our Building Blocks}\label{sec:combine-Combining DP Mechanism with Our Building Blocks}
In this sub-section we present our solution of combining DP mechanism with building blocks,
that is,
how to train a ML classifier on the mixed dataset that combine an encrypted dataset and a noise dataset.

\begin{figure}[htb]
\centering
\includegraphics[width=8.5cm]{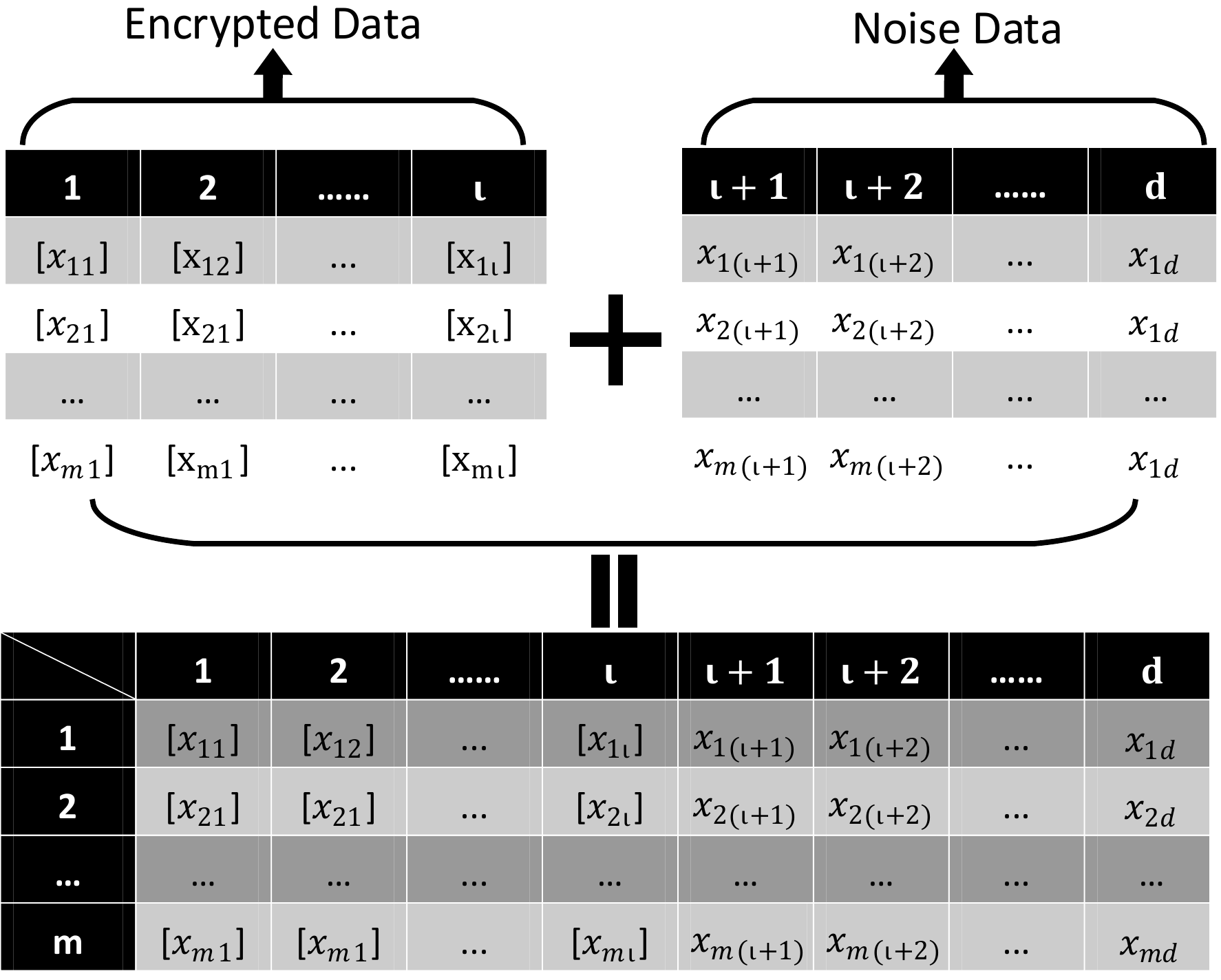}
\caption{A Mixed Dataset: the Encrypted Dataset Combine with the Noise Dataset.}\label{fig:A Mixed Dataset: Encrypted Data Combine with Noise Data}
\end{figure}

The idea is simple but effective:
We still let the high scores part has $\iota$ features.
After requesting data from $\mathcal{P}$, $\mathcal{U}$ obtains a dataset mixed with the encrypted dataset and the noise dataset, as depict in Figure \ref{fig:A Mixed Dataset: Encrypted Data Combine with Noise Data}.
In the process of iteratively updating parameters,
learning rate in the high scores part is computed by Algorithm \ref{algorithm:Training Private-Preserving LR},
and learning rate in the low scores part is computed by Formula \ref{equ:iteration formula of LR} just as in normal plaintext case.

Here, we present a simplified example to understand:
Suppose we are computing $w=X+Y$.
$X= \left\{ {\left[ {{x}_{1}}\right],\left[{{x}_{2}}\right],\ldots,
}\right.\\ \left.{
\left[{{x}_{ \iota }} \right],{{x}_{ \iota +1}},{{x}_{ \iota +2}},\ldots ,
{{x}_{d}}} \right\}$,
$ Y =\left\{{ \left[ {{y}_{1}}\left], \right[{{y}_{2}}\left],\ldots, \right[{{y}_{ \iota }} \right],{{y}_{ \iota +1}},
}\right.\\ \left.{
{{y}_{ \iota +2}},\ldots,{{y}_{d}} }\right\}$.
Then, $w = \left\{{ [ {{x}_{1}}+{{y}_{1}}] , [{{x}_{2}}+{{y}_{2}}] ,\ldots , [{{x}_{ \iota }}+{{y}_{ \iota }}]
}\right.\\ \left.{
,{{x}_{ \iota +1}}+{{y}_{ \iota +1}},{x}_{ \iota +2}+{{y}_{ \iota +2}},\ldots ,{{x}_{d}}+{{y}_{d}}} \right\}$.
Analogous, when the training algorithm updates parameters, the learning rate of the encrypted dataset is computed by Algorithm \ref{algorithm:Training Private-Preserving LR}, and the learning rate of the noise dataset is computed as normal.

\section{Security Analysis}\label{sec:Security Analysis}

When facing the honest-but-curious adversaries,
we follow a commonly used definition -- secure two-party computation (cf. Appendix \ref{sec:appendix-Secure two-party computation}) and a useful theorem -- modular sequential composition (cf. Appendix \ref{sec:appendix-Modular sequential composition}).
We present our security proofs according to the ideas of these two definitions.
%Because of the layout restrictions,
%we don't elaborate on these two definitions.
For more details, we refer the reader to \cite{10} for secure two-party computation and \cite{15} for modular sequential composition.

\begin{myproof}[Proof of Proposition \ref{proposition:The security of building blocks}]\label{myproof:Proposition 3}
\ \\\indent
\textit{Security proof for Protocol \ref{protocol:Secure Addition Protocol}, \ref{protocol:Secure Subtraction Protocol}, \ref{protocol:Secure dot product} and \ref{protocol:Secure dot product 2}:}
Protocol \ref{protocol:Secure Addition Protocol}, \ref{protocol:Secure Subtraction Protocol} and \ref{protocol:Secure dot product} construct secure two-party computation using Paillier's additively homomorphic property.
Protocol \ref{protocol:Secure dot product 2} constructs secure two-party computation using RSA's multiplicative homomorphic property.
In these four protocols, Alice does not receive any message whose view only consists in its input,
and Alice does not call to any other protocols.
%We symbolize the output as ${{\left[ \text{v} \right]}_{A}}$.
The security proofs of these four protocols can be summarized as follows.
\\\indent
The input of Alice is $\left( a,{\sf{PK}_{A}},{\sf{SK}_{A}} \right)$, Bob is $\left({ b\ or\ {{\left[ b \right]}_{A}},
}\right.\\ \left.{
{\sf{PK}_{A}} }\right)$.
\\\indent
As Alice does not receive any message, Alice's view only consists in its input.
Hence the simulator $S_{A}^{\pi }\left( a,F\left( a,b \right) \right)=view_{A}^{\pi }\left( a,{\sf{PK}_{A}},{\sf{SK}_{A}} \right)$.
\\\indent
Bob's view is $view_{B}^{\pi }\left(b\ or\ {{\left[ b \right]}_{A}},{\sf{PK}_{A}},{{[a]}_{A}}, {Output_{B}} \right)$.
$a$ is encrypted by ${\sf{PK}_{A}}$,
and the confidentiality of ${[a]}_{A}$ is equivalent to the cryptosystem,
thereby Bob cannot infer the value of ${[a]}_{A}$ straightforward.
%, Bob can't infer is value straightforward.
The simulator $S_{B}^{\pi }$ does the following:
(i) Generates $l$ random coins, obtains ${[c]_{A}}={{[ \{ {{c}_{1}},{{c}_{2}},\ldots ,{{c}_{l}} \} ]}_{A}}$ by $P{{K}_{A}}$, where $l$ is the length of a.
(ii) Outputs $\left( {{[c]}_{A}},{\sf{PK}_{A}},b\ or\ {{[b]}_{A}},{[f(c,b)]}_{A}  \right)$.
By semantic security of the used cryptosystem:
$\left( {{[c]}_{A}},{\sf{PK}_{A}},b\ or \ {{\left[ b \right]}_{A}},{{\left[ F\left( c,b \right) \right]}_{A}} \right) \approx \left( {{[ a ]}_{A}},
{\sf{PK}_{A}},b\ or\ {{\left[ b \right]}_{A}},{{[F(a,b)]}_{A}} \right)$.

\textit{Security proof for Protocol \ref{protocol:Security exponential Power}:}
The input of Alice is $\left( a,{\sf{PK}_{A}},{\sf{SK}_{A}} \right)$, Bob is $\left( b,{\sf{PK}_{A}} \right)$.
\\\indent
As Alice does not receive any message in this hybrid model, Alice's view only consists in its input.
Hence the simulator $S_{A}^{\pi }\left( a,F\left( a,b \right) \right)=view_{A}^{\pi }\left( a,{\sf{PK}_{A}},{\sf{SK}_{A}} \right)$.
\\\indent
Bob's view is $view_{B}^{\pi }\left( {b,{\sf{PK}_{A}},{{||{e}^{a}||}_{A}},{{F}_{\text{protocol}-4}}({||{e}^{a}||}_{A},
}\right.\\ \left.{
{||b||}_{A} ),{Output_{B}}} \right)$.
${{e}^{a}}$ is encrypted by ${\sf{PK}_{A}}$,
and the confidentiality of ${||{e}^{a}||}_{A}$ is equivalent to the cryptosystem,
So Bob cannot infer its value straightforward.
The simulation $S_{B}^{\pi }$, on input $( b,\sf{PK}_{A} )$, dose the following:
(i) Generates $l$ random coins, then obtains $c=\left\{ {{c}_{1}},{{c}_{2}},\ldots ,{{c}_{l}} \right\}$ and ${||{e}^{c}||}_{A}$ by ${\sf{PK}_{A}}$, where $l$ is the length of $a$.
(ii) Outputs $\left({ c,{\sf{PK}_{A}},{\sf{SK}_{A}},{{{||{e}^{cb}||}}_{A}},
}\right.\\ \left.{
{{F}_{\text{protocol}-4}}\left( {||{e}^{c}||}_{A},{||b||}_{A}  \right) }\right)$.
As RSA is semantically secure, the distributions $S_{B}^{\pi }=\left( b,{\sf{PK}_{A}},{ ||{e}^{c}||}_{A},{||{e}^{cb}||}_{A} \right)$
and $view_{B}^{\pi }=\left( b,{\sf{PK}_{A}},{{||{e}^{a}||}_{A}},{{||{e}^{ab}||}_{A}} \right)$ are computationally indistinguishable.
\\\indent
As Protocol \ref{protocol:Secure dot product 2} is secure in the honest-but-curious model, we obtain the security of the Protocol 5 using Theorem \ref{theorem:Modular sequential composition}.

\textit{Security proof for Protocol \ref{protocol:Converting Ciphertext 1}:}
The function is $F$: $F\left( {||{e}^{ab}||}_{A},{\sf{PK}_{P}},{\sf{SK}_{P}},{\sf{PK}_{R}},{\sf{SK}_{R}} \right)=\left( \phi,{[[{e}^{ab}]]}_{A} \right)$.
\\\indent
Alice's view is $view_{A}^{\pi }=\left({ {[[{e}^{ab+r}]]}_{A},{\sf{PK}_{P}},{\sf{SK}_{P}},{\sf{PK}_{R}},
 }\right.\\ \left.{
{\sf{SK}_{R}},{output_{Protocal-3}},  }\right)$,
and ${output_{Protocal-3}}={||{e}^{ab+r}||}_{A}$.
$S_{A}^{\pi }$ runs as follows:
(i) Generates $l$ random coins, then obtains $c=\left\{ {{c}_{1}},{{c}_{2}},\ldots ,{{c}_{l}} \right\}$ and ${[[ {e}^{c} ]]}_{A}$ by ${\sf{PK}_{P}}$, where $l$ is the length of $\left( ab+r \right)$.
(ii) Outputs $\left( {{||{e}^{c}||}_{A}, {[[{e}^{c}]]}_{A},{\sf{PK}_{P}},{\sf{SK}_{P}}, {\sf{PK}_{R}},{\sf{SK}_{R}}
  }\right.\\ \left.{
  }\right)$.
$(ab+r)$ and $c$ are taken from the same distribution, independently from any other parameter.
Paillier and RSA is semantically secure.
So $\left(  {||{e}^{c}||}_{A},{[[{e}^{c}]]}_{A},{\sf{PK}_{P}},{\sf{SK}_{P}},{\sf{PK}_{R}},{\sf{SK}_{R}} \right)\approx\\
\left( {||{e}^{ab+r}||}_{A},{[[{e}^{ab+r}]]}_{A},{\sf{PK}_{P}},{\sf{SK}_{P}},{\sf{PK}_{R}},{\sf{SK}_{R}} \right)$.
\\\indent
Bob's view is $view_{B}^{\pi }=\left( {||{e}^{ab}||}_{A},{||{e}^{ab+r}||}_{A},r,{[[{e}^{ab+r}]]}_{A}  \right)$, $S_{B}^{\pi }$ runs as follows:
(i) Generates $l$ random coins, obtains $c=\left\{ {{c}_{1}},{{c}_{2}},\ldots ,{{c}_{l}} \right\}$, where $l$ is the length of $ab$.
(ii) obtains ${[{e}^{c}]}_{A}$ and ${[{e}^{c+r}]}_{A}$ by ${\sf{PK}_{P}}$.
(ii) Outputs $\left( {{||{e}^{c}||}_{A} ,{||{e}^{c+r}||}_{A},{[[{e}^{c+r}]]}_{A},
}\right.\\ \left.{
r,{\sf{PK}_{P}},{\sf{SK}_{P}},{\sf{PK}_{R}},{\sf{SK}_{R}} }\right)$.
The distribution of $c$ and $ab$ are identical and RSA is semantically secure, so the real distribution $\left\{ r,{||{e}^{ab}||}_{A} \right\}$ and the ideal distribution $\left\{  r, {||{e}^{c}||}_{A}\right\}$ are statistically indistinguishable.
\\\indent
As Protocol \ref{protocol:Secure dot product} is secure in the honest-but-curious model, we obtain the security of the Protocol \ref{protocol:Converting Ciphertext 1} using Theorem \ref{theorem:Modular sequential composition}.

\textit{Security proof for Protocol \ref{protocol:Converting Ciphertext 2}:}
The function is $F$: $F\left({ {[[b]]}_{A},
}\right.\\ \left.{
{\sf{PK}_{A}},{\sf{SK}_{A}},{\sf{PK}_{B}},{\sf{SK}_{B}} }\right)= ( \phi ,{{[[b]]}_{B}} )$.
\\\indent
Alice's view is $view_{A}^{\pi }=\left( {{ [[b+r ]]}_{A}},{\sf{PK}_{A}},{\sf{SK}_{A}},{\sf{PK}_{B}} \right)$.
$S_{A}^{\pi }$ runs as follows:
(i) Generates $l$ random coins and obtains ${{{[[c]]}_{A}}={{[[ \left\{ {{c}_{1}},{{c}_{2}},\ldots ,{{c}_{l}} \right\} ]]}_{A}}}$ by ${\sf{PK}_{A}}$, where $l$ is the length of $( m+r )$.
(ii) Outputs $\left( {[[ c ]]}_{B},{[[ c ]]}_{A},{\sf{PK}_{A}},{\sf{SK}_{A}},{\sf{PK}_{B}} \right)$.
$(b+r)$\\
and $c$ are taken from the same distribution, independently from any other parameter, and Paillier is semantically secure,
so $\left( {[[ c ]]}_{B},{[[ c ]]}_{A},{\sf{PK}_{P}},{\sf{SK}_{P}},{\sf{PK}_{R}},{\sf{SK}_{R}}\right)\approx
\left({  {[[ b+r ]]}_{B},{[[ b+r ]]}_{A},
}\right.\\ \left.{
{\sf{PK}_{P}},{\sf{SK}_{P}},{\sf{PK}_{R}},{\sf{SK}_{R}} }\right)$.
\\\indent
Bob's view is $view_{B}^{\pi }=\left({ {{[[b]]}_{A}},{{[[ b+r ]]}_{A}},r,{\sf{PK}_{B}},{\sf{SK}_{B}},
}\right.\\ \left.{
{\sf{PK}_{A}} }\right)$, $S_{B}^{\pi }$ runs as follows:
(i) Generates $l$ random coins and obtains ${{[[ c ]]}_{A}}={{[[ \left\{ {{c}_{1}},{{c}_{2}},\ldots ,{{c}_{l}} \right\} ]]}_{A}}$ by ${\sf{PK}_{A}}$, where $l$ is the length of $m$.
(ii) Outputs $\left({ {{[[ c ]]}_{B}},{{[[ c+r ]]]}_{A}},r,{\sf{PK}_{B}},{\sf{SK}_{B}},{\sf{PK}_{A}}
}\right.\\ \left.{
 }\right)$.
The distribution of $c$ and $m$ are identical, and Paillier is semantically secure,
so the real distribution $\left\{{ {{[[ m ]]}_{B}},[[m+r
}\right.\\ \left.{
{{  ]]}_{A}},r;{{[[ m+r ]]}_{B}} }\right\}$ and the ideal distribution $\left\{{ {{[[ c ]]}_{B}},[[ c+r ]]
}\right.\\ \left.{
{{ }_{A}},r;{{[[ c+r ]]}_{B}} }\right\}$ are statistically indistinguishable.
\\\indent
As Protocol \ref{protocol:Secure Addition Protocol} is secure in the honest-but-curious model, we obtain the security of the Protocol \ref{protocol:Converting Ciphertext 2} using Theorem \ref{theorem:Modular sequential composition}.
$\hfill\square$
\end{myproof}

\begin{myproof}[Proof of Proposition \ref{proposition:Training Private-Preserving LR}]\label{myproof:algorithm-Training Private-Preserving LR}
%$\mathcal{U}$ and ${\mathcal{P}_{i}}$ have 3 interactions with each other ().
Since each ${\mathcal{P}_{i}}$ behaves in the same way, we use ${\mathcal{P}_{i}}$ substitute each $\mathcal{P}$'s behavior in this security proof.
\\\indent
$\mathcal{U}$'s view is $view_{\mathcal{U}}^{\pi }=\left({ {\sf{PK}_{\mathcal{U}}},{\sf{SK}_{\mathcal{U}}},{{[[{e}^{X}]]}_{{\mathcal{P}_{i}}}},
{||{e}^{X}||}_{{\mathcal{P}_{i}}},
}\right.\\ \left.{
{[[ {e}^{X\beta}]]}_{{\mathcal{P}_{i}}},{[[ \frac{X}{{e}^{\beta X+r}+{{e}^{r}}} ]]}_{{\mathcal{P}}_{i}},
{[[\underset{i=1}{\overset{m}{\mathop \sum }}X\frac{{{e}^{\beta X}}}{1+{e}^{\beta X}}]]}_{{\mathcal{P}}_{i}},
{Output}_{\mathcal{U}}}\right)$.
As intermediate results are encrypted by the public key of each $\mathcal{P}_{i}$, and Paillier and RSA is semantically secure, thus the sensitive information of each $\mathcal{P}_{i}$ is computationally indistinguishable in intermediate results.
What we need to discuss is the confidentiality of ${Output}_{\mathcal{U}}=\left( \beta,\underset{i=1}{\overset{m}{\mathop \sum }}\left( X\frac{{{e}^{ \beta X}}}{1+{{e}^{ \beta X}}}-XY \right)\right)$,
that is, whether $\mathcal{U}$ can guess the sensitive information of each $\mathcal{P}_{i}$ from ${Output}_{\mathcal{U}}$ successfully.
\\\indent
Obviously $a=\underset{i=1}{\overset{m}{\mathop \sum }}\left( X\frac{{{e}^{ \beta X}}}{1+{{e}^{ \beta X}}}-XY \right)$ is a no-solution equation for the unknown x and the known $\beta$.
In addition to brute force cracking, there is no other better way to get the value of $X$.
We assume a small dataset has 2-dimensional 100 instances, and the length of each dimension is 32 bits\footnote{Typically, single-precision floating-point occupies 4 bytes (32-bit) memory space.}.
Then the probability that $\mathcal{U}$ guesses success is $\frac{1}{2^{32\times d\times m}}=\frac{1}{2^{6400}}$.
It is a negligible probability of success \cite{1}.
\\\indent
${\mathcal{P}_{i}}$'s view is $view_{{\mathcal{P}_{i}}}^{\pi }=\left({ {\sf{PK}_{{\mathcal{P}_{i}}}},{\sf{SK}_{{\mathcal{P}_{i}}}},{{\left[\left[ {{e}^{ \beta X+r}}+{{e}^{r}}\right]\right]}_{{\mathcal{P}_{i}}}},
}\right.\\ \left.{
{||{e}^{X}||}_{{\mathcal{P}_{i}}},{{[[{e}^{X}]]}_{{\mathcal{P}_{i}}}} }\right)$.
${\mathcal{P}_{i}}$ does not output any message.
Hence the simulator $S_{{\mathcal{P}_{i}}}^{\pi }=view_{{\mathcal{P}_{i}}}^{\pi }$.
$P_{i}$ runs as follows:
(i) Generates $l$ random coins and obtains $c=\left\{ {{c}_{1}},{{c}_{2}},\ldots ,{{c}_{l}} \right\}$, where $l$ is the length of $r$.
(ii) Uniformly picking $m=\left\{ {{m}_{1}},{{m}_{2}},\ldots ,{{m}_{d}} \right\}$, where $m\in {{\mathbb{M}}_{{\mathcal{P}_{i}}}}$.
(iii) Output $\left( {\sf{PK}_{{\mathcal{P}_{i}}}},
{\sf{SK}_{{\mathcal{P}_{i}}}},{{\left[\left[ {{e}^{mX+c}}+{{e}^{c}} \right]\right]}_{{\mathcal{P}_{i}}}} \right)$.
The distribution of $\left( c,m \right)$ and $\left( r, \beta  \right)$ are identical,
so the real distribution $\left({ {\sf{PK}_{{\mathcal{P}_{i}}}},{\sf{SK}_{{\mathcal{P}_{i}}}},
}\right.\\ \left.{
{{\left[\left[ {{e}^{ \beta X+r}}+{{e}^{r}} \right]\right]}_{{\mathcal{P}_{i}}}} }\right)$
and the ideal distribution $\left({ {\sf{PK}_{{\mathcal{P}_{i}}}},{\sf{SK}_{{\mathcal{P}_{i}}}},
}\right.\\ \left.{
 {{[[ {{e}^{mX+c}}+{{e}^{c}} ]]}_{{\mathcal{P}_{i}}}} }\right)$ are statistically indistinguishable.
\\\indent
As those Protocols used in Algorithm \ref{algorithm:Training Private-Preserving LR} are secure in the honest-but-curious model, we obtain the security of Algorithm \ref{algorithm:Training Private-Preserving LR} using Theorem \ref{theorem:Modular sequential composition}.
$\hfill\square$
\end{myproof}

\begin{myproof}[Proof of Proposition \ref{proposition:the security of heda}]\label{myproof:algorithm-the security of heda}
\ \\\indent
Each $\mathcal{P}$ computes feature scores by feature evaluation technologies locally,
obtains the noise dataset by Algorithm \ref{algorithm:IMA e-DP Mechanism},
and interacts with $\mathcal{U}$ to process the encrypted dataset by Algorithm \ref{algorithm:Training Private-Preserving LR}.
$\mathcal{U}$ trains a LR classifier with the encrypted datasets by Algorithm \ref{algorithm:Training Private-Preserving LR}.
\\\indent
As those Protocols or Algorithms used in Algorithm \ref{algorithm:Privacy-Preserving Training} are secure in the honest-but-curious model, we obtain the security using Theorem \ref{theorem:Modular sequential composition}.
$\hfill\square$
\end{myproof}

\section{Performance Evaluation}\label{sec:PE}
We present the evaluation of \texttt{Heda} in this section.
We answer the following questions in our evaluations toward \texttt{Heda}:
(i) the performance of our DP components,
(ii) the performance of our building blocks,
and (iii) the performance overhead of \texttt{Heda}.
\subsection{Preparations}

\subsubsection{Implementations}\label{sec:PE-Implementations}
Our experiments were run using a desktop computer with configuration: single Intel i7 (i7-3770 64bit) processor for total 4 cores running at 3.40GHz and 8 GB RAM.
We have implemented the building blocks, DP components, \texttt{Heda}, and mRMR in Java Development Kit 1.8.
Feature selection algorithms including Chi-square test, Pearson correlation, Spearman correlation, Random forest and KW have been implemented in scikit-learn\footnote{http://scikit-learn.org}.
%We have implemented Kruskal-Wallis H in Weka \cite{51}.

The operations of Paillier and RSA are carried out in finite fields involving modular arithmetic operations on integers, while classifiers training generally use floating point numbers.
In order to encrypted data taking real values, it is necessary to previously perform a format conversion into an integer representation.
According to the international standard IEEE 754, binary floating point number $D$ is expressed as $D={{\left( -1 \right)}^{s}}\times M\times {{2}^{E}}$,
where $s$ is the sign bit, $M$ is a significant number, and $E$ is the exponent bit.
We employ it to perform the format conversion toward our implementations.
Overlength effective bits lead to inefficient algorithms,
while underlength effective bits may low the accuracy.
We retained two decimal places empirically.
%and experiments show that minor precision loss in the intermediate results doesn't affect the accuracy of the final model.
 %(cf. Section \ref{sec:PE-Building Blocks Performance}).

Plaintext should be guaranteed in the plaintext space of the used cryptosystem.
So we must consider the key length to avoid the possibility of overflow.
%Algorithm \ref{algorithm:Training Private-Preserving LR} includes operations such as seure addition, secure dot product, and secure power exponent.
After analyzing all the intermediate results, Paillier's key N is set to 2048-bit and RSA's key N is 2048-bit.

\subsubsection{Datasets}
Four datasets from the UCI ML repository \cite{52} originated from several different application domains are used in our evaluation:
(i)	Breast Cancer Wisconsin (Diagnostic) Dataset (BCWD),
(ii) Adult Dataset (Adult),
(iii) Credit Approval Dataset (CAD),
(iv) Car Evaluation Dataset (Car).
The statistics are shown in Table \ref{table:Statistics of Datasets}.
In order to avoid overfitting or contingent results, we show the average results of cross-validation of 10 runs.
In each cross-validation, we randomly take 80\% to train the model, and the remainder for testing.

\begin{table}[!t]
\normalsize
\centering
\small
\caption{Statistics of Datasets}\label{table:Statistics of Datasets}
\renewcommand{\arraystretch}{1.1}
\begin{tabular}{|p{1.2cm}<{\centering}|p{1.4cm}<{\centering}|p{1.6cm}<{\centering}|p{1.3cm}<{\centering}|p{1.3cm}<{\centering}|}
\hline
Datasets	&Instances number&	Attributes number&	Discrete attributes &	Numerical attributes \\
\hline
\hline
BCWD&	699	&9&	0&	9\\
\hline
Adult&	32561&	14&	8&	6\\
\hline
CAD	&690	&15&	9&	6\\
\hline
Car	&1728&	6&	0&	6\\
\hline
\end{tabular}
\end{table}

\subsection{Evaluation of Differential Privacy Components}\label{sec:PE-DP Components Performance}
%We have explored the balance between  usability and confidentiality in DP mechanism.
In DP mechanism, we employ IMA to reduce the sensitivity $\Delta f$ and give a formula to determining the appropriate privacy budget $\epsilon $.
We recall that the best cluster size is $k=\sqrt[{}]{\frac{m}{2}}$, i.e. the best cluster size of the four datasets -- BCWD, Adult, CAD and Car -- are 16, 127, 18 and 29 respectively.
Standard DP (cf. Definition \ref{def:Laplace mechanism}) is the baseline for comparison, where the $\epsilon $ for each dataset is obtained by Algorithm \ref{algorithm:generating the appropriate value of e}.
The scheme proposed by Soria et al. \cite{17} is employed to serve as a control group named IMDAV, which is the state of the art for reducing the sensitivity in DP mechanisms based on IMA.
Sum of Squared Errors (SSE) is used to evaluate the information loss of the dataset,
and the percentage of Record Linkages (RL) is used to evaluate the risk of privacy leak.
%which are the general evaluation criteria employed by the k-anonymity research community \cite{17,53}.

\begin{figure*}
\begin{minipage}[t]{0.25\textwidth}
    \centering
    \includegraphics[height=3.5cm]{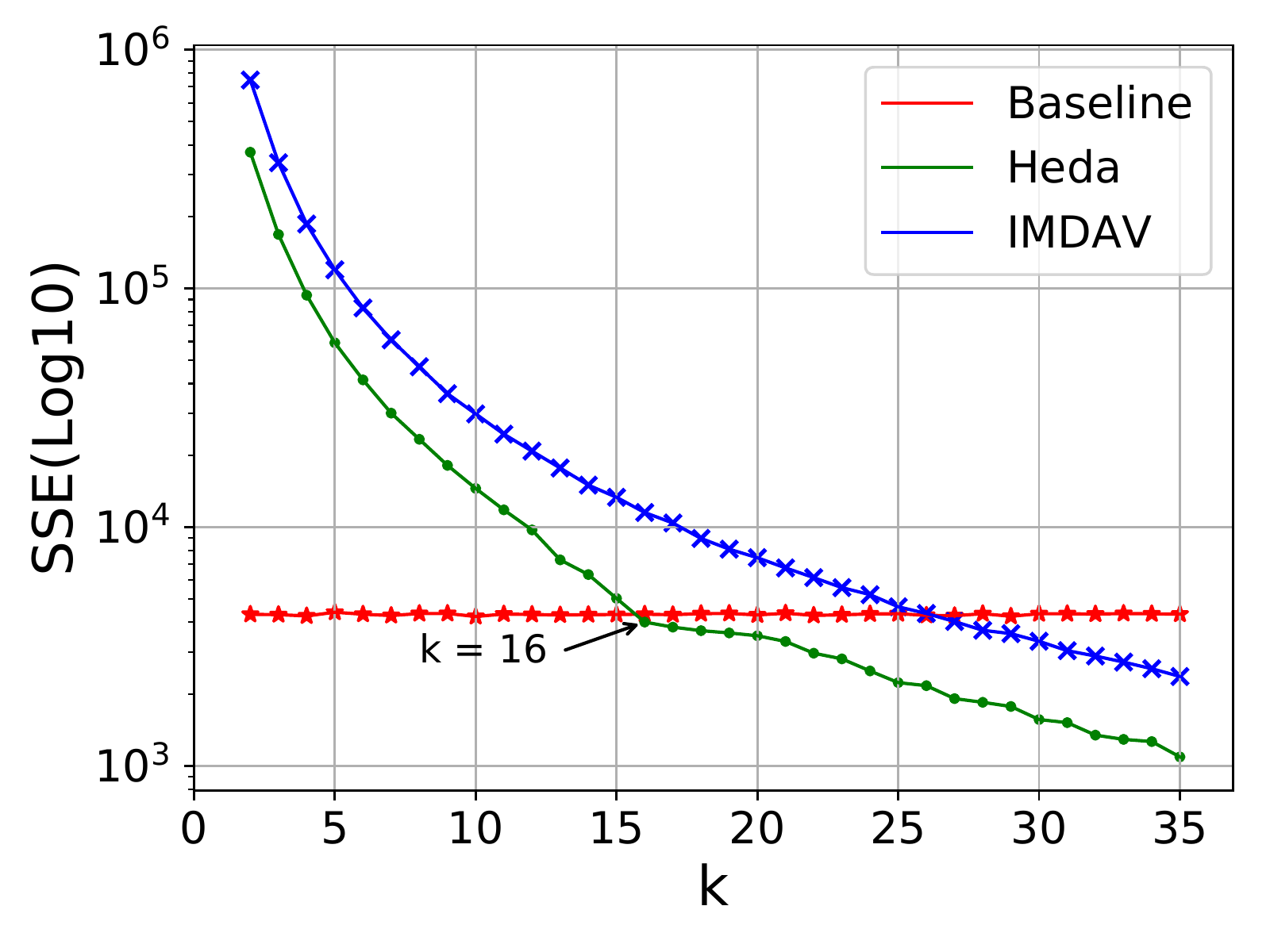}\\
    \scriptsize{(a) BCWD}
\end{minipage}%
\begin{minipage}[t]{0.25\textwidth}
    \centering
    \includegraphics[height=3.5cm]{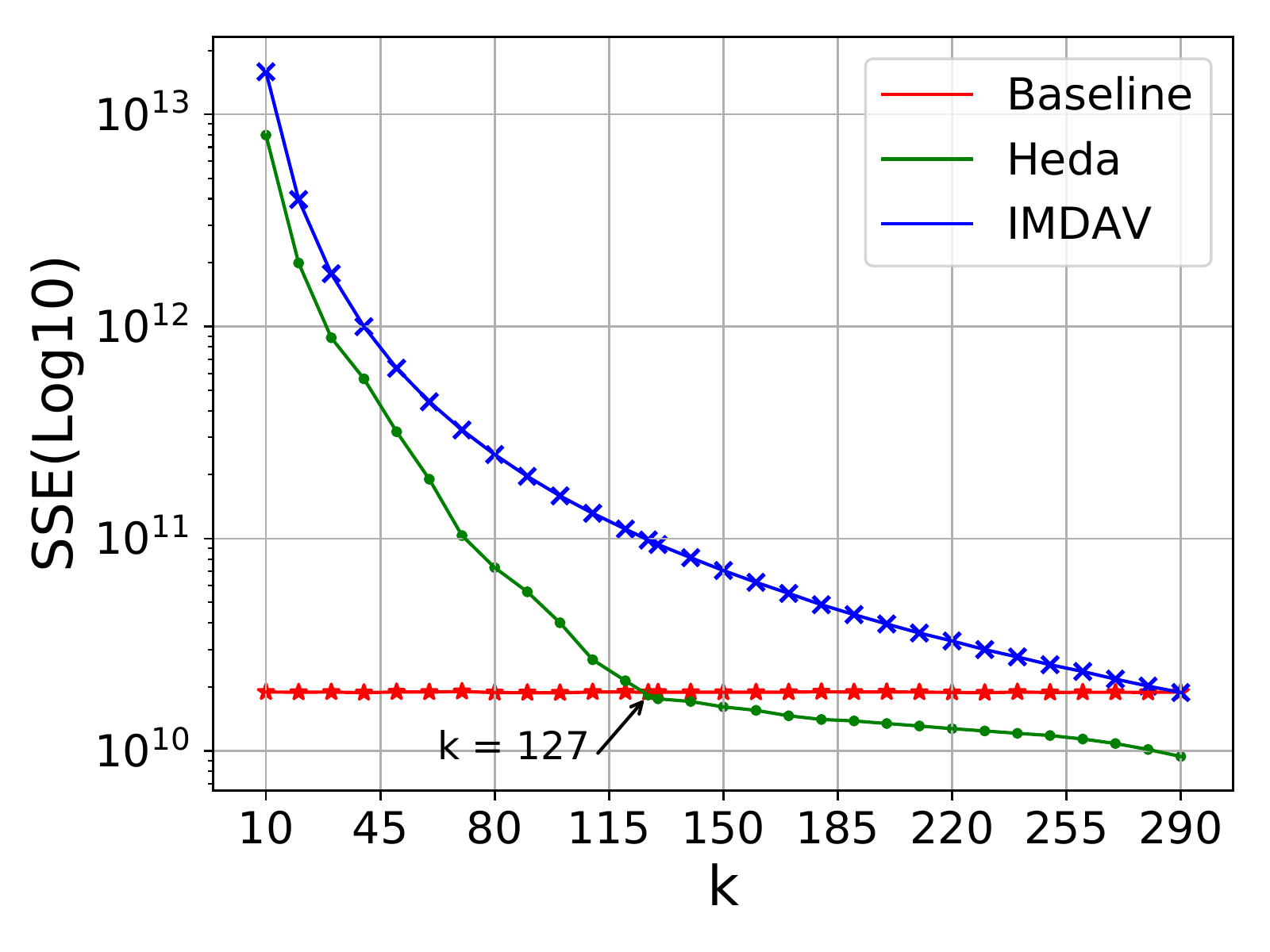}\\
    \scriptsize{(b) Adult}
\end{minipage}%
\begin{minipage}[t]{0.25\textwidth}
    \centering
    \includegraphics[height=3.5cm]{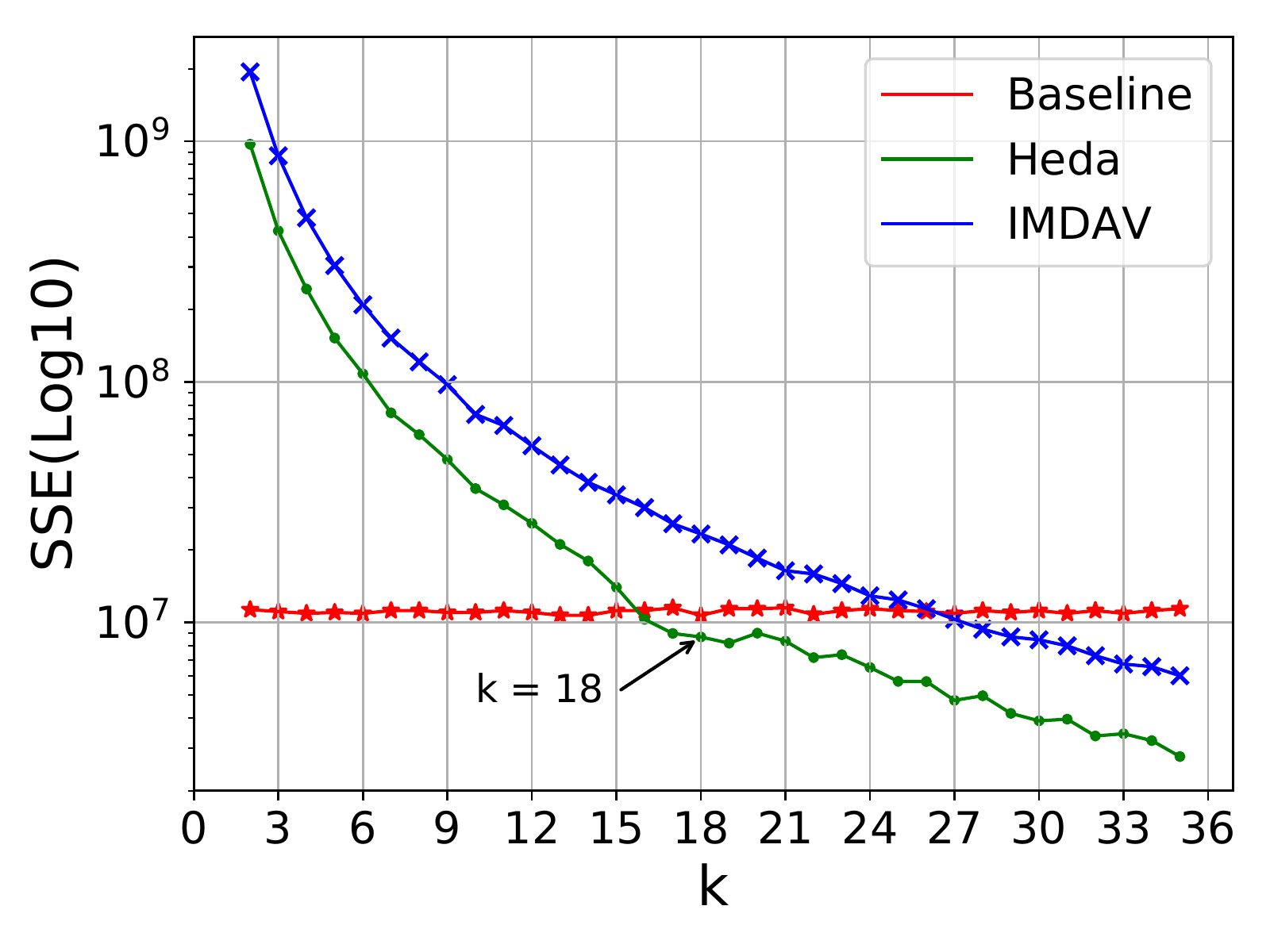}\\
    \scriptsize{(c) CAD}
\end{minipage}%
\begin{minipage}[t]{0.25\textwidth}
    \centering
    \includegraphics[height=3.5cm]{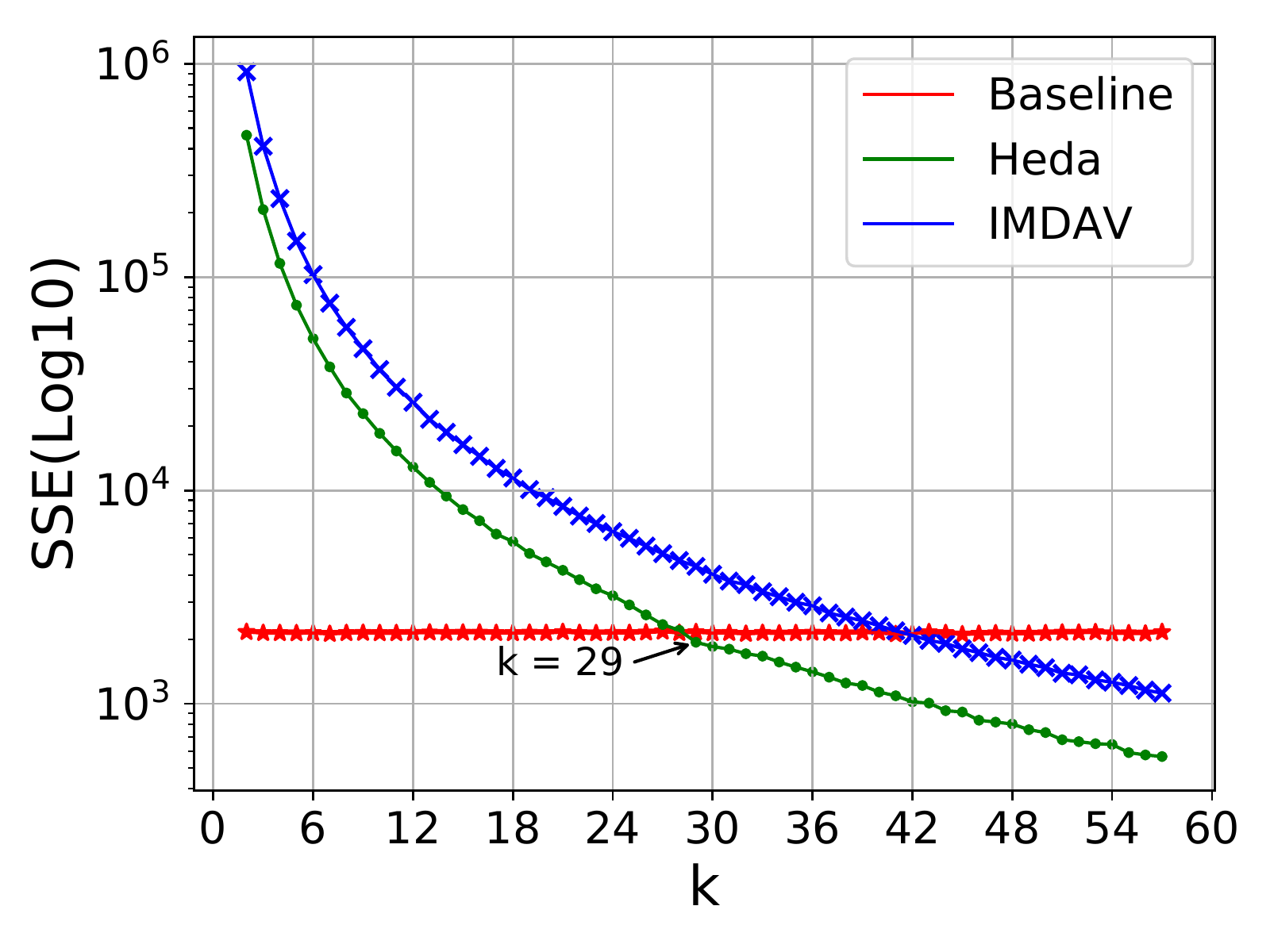}\\
    \scriptsize{(d) Car}
\end{minipage}%
\caption{Differential Privacy Components Performance: Sum of Squared Errors.}
\label{fig:Privacy Differential Components Performance: Sum of Squared Errors}
\end{figure*}

For a given dataset $D$ and an $\epsilon $-DP dataset ${{D}_{\epsilon }}$ with microaggregation of $k$ size, SSE (Formula \ref{equ:SSE}) is defined as the sum of squares of attribute distances between original records in $D$ and their versions in the $\epsilon $-DP dataset ${{D}_{\epsilon }}$.
\begin{equation}\label{equ:SSE}
  SSE=\underset{{{A}_{j}}\in D,A_{j}^{'}\in {D}'}{\mathop \sum }\,\underset{{{x}_{ij}}\in {{A}_{j}},x_{ij}^{'}\in A_{j}^{'}}{\mathop \sum }\,{{\left( {{x}_{ij}}-x_{ij}^{'} \right)}^{2}}
\end{equation}
We implement our DP components (Algorithm \ref{algorithm:IMA e-DP Mechanism}), Standard DP and IMDAV on the four datasets with different cluster sizes, then record different SSE values.
The results are depicted in Figure \ref{fig:Privacy Differential Components Performance: Sum of Squared Errors}.
It is clear that our DP components have low information loss compare to IMDAV.
When $k = \sqrt[{}]{\frac{m}{2}}$, our DP components have lower SSE than Baseline.

\begin{figure*}
\begin{minipage}[t]{0.25\textwidth}
    \centering
    \includegraphics[height=3.5cm]{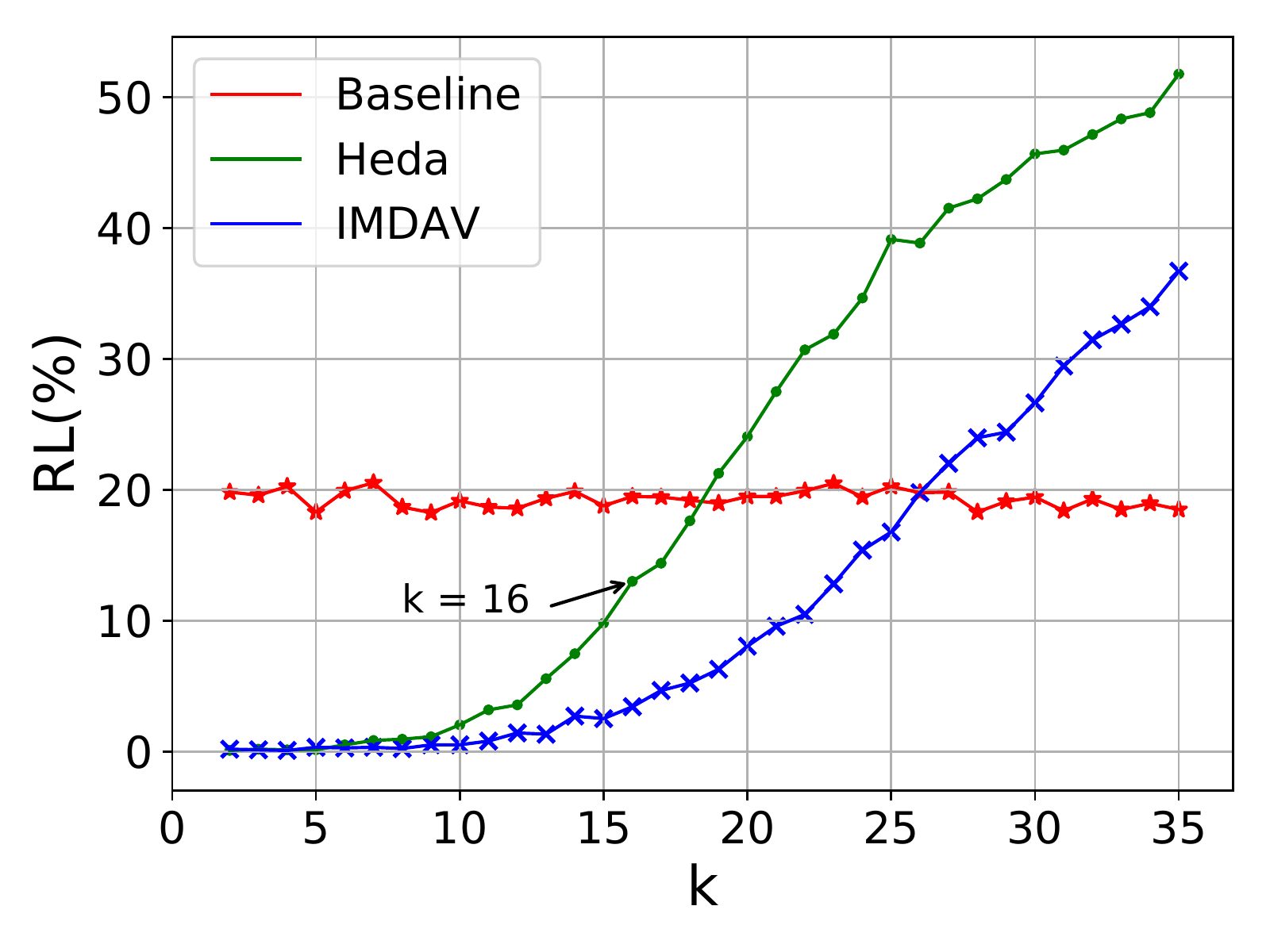}\\
    \scriptsize{(a) BCWD}
\end{minipage}%
\begin{minipage}[t]{0.25\textwidth}
    \centering
    \includegraphics[height=3.5cm]{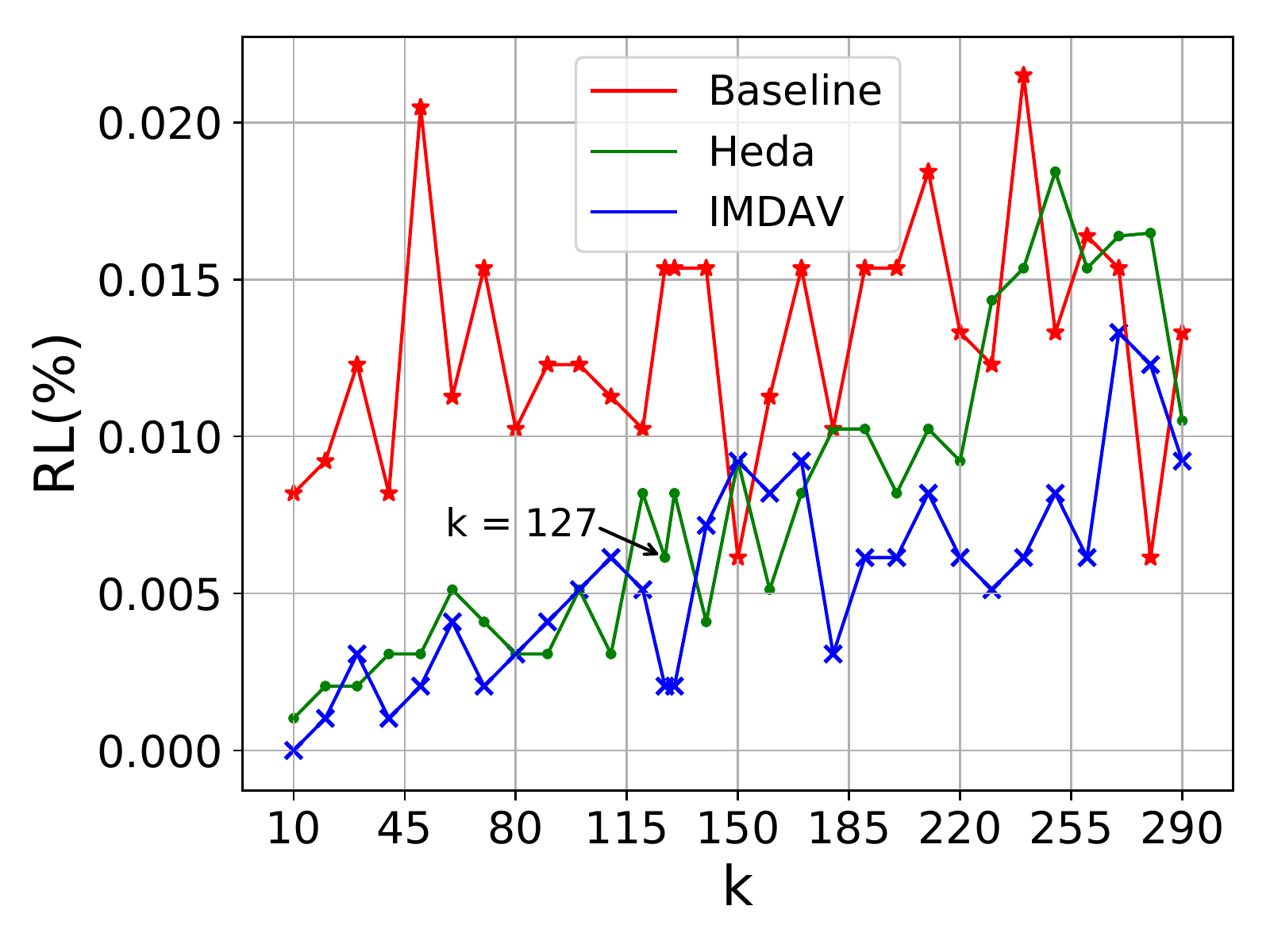}\\
    \scriptsize{(b) Adult}
\end{minipage}%
\begin{minipage}[t]{0.25\textwidth}
    \centering
    \includegraphics[height=3.5cm]{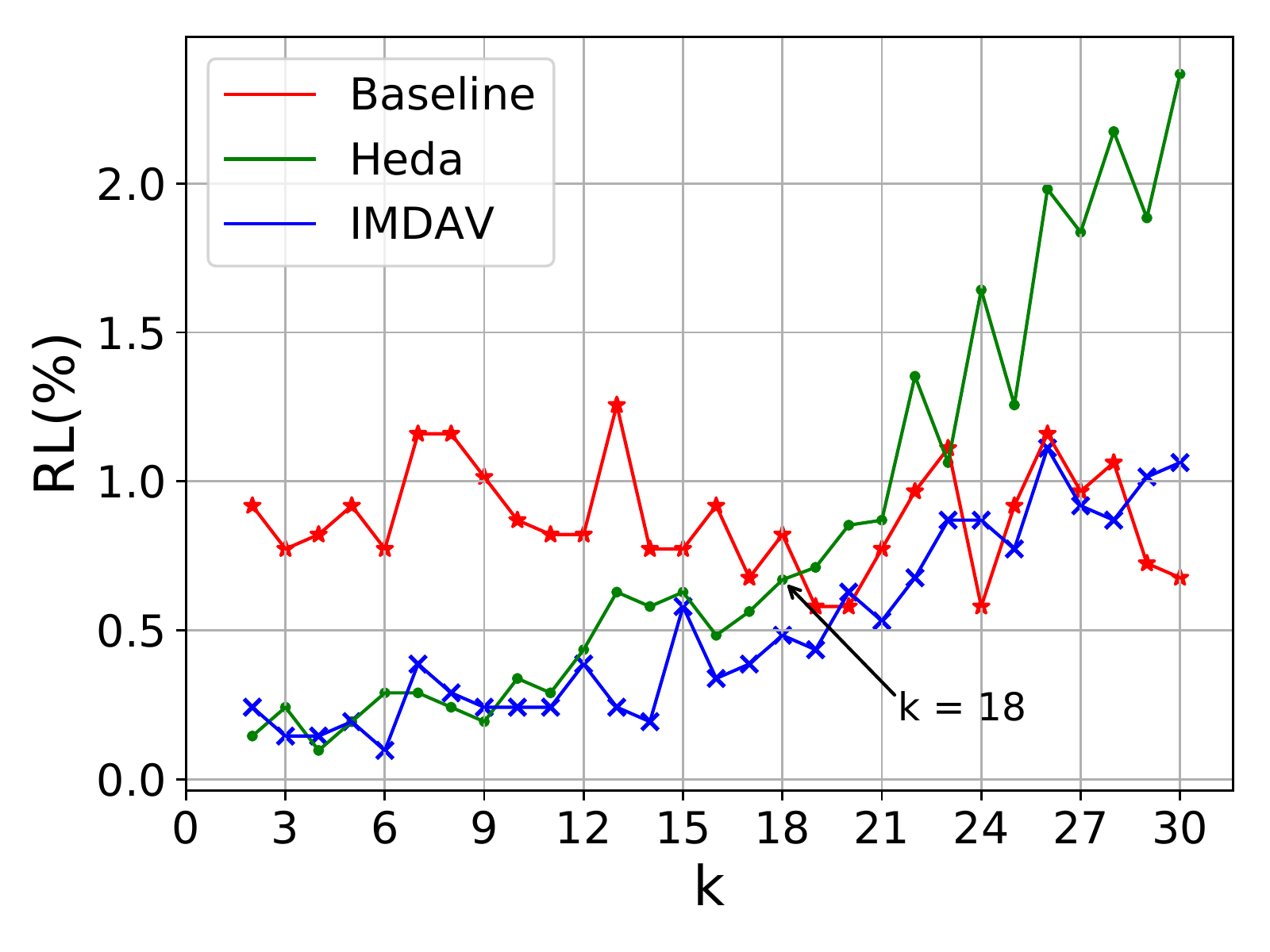}\\
    \scriptsize{(c) CAD}
\end{minipage}%
\begin{minipage}[t]{0.25\textwidth}
    \centering
    \includegraphics[height=3.5cm]{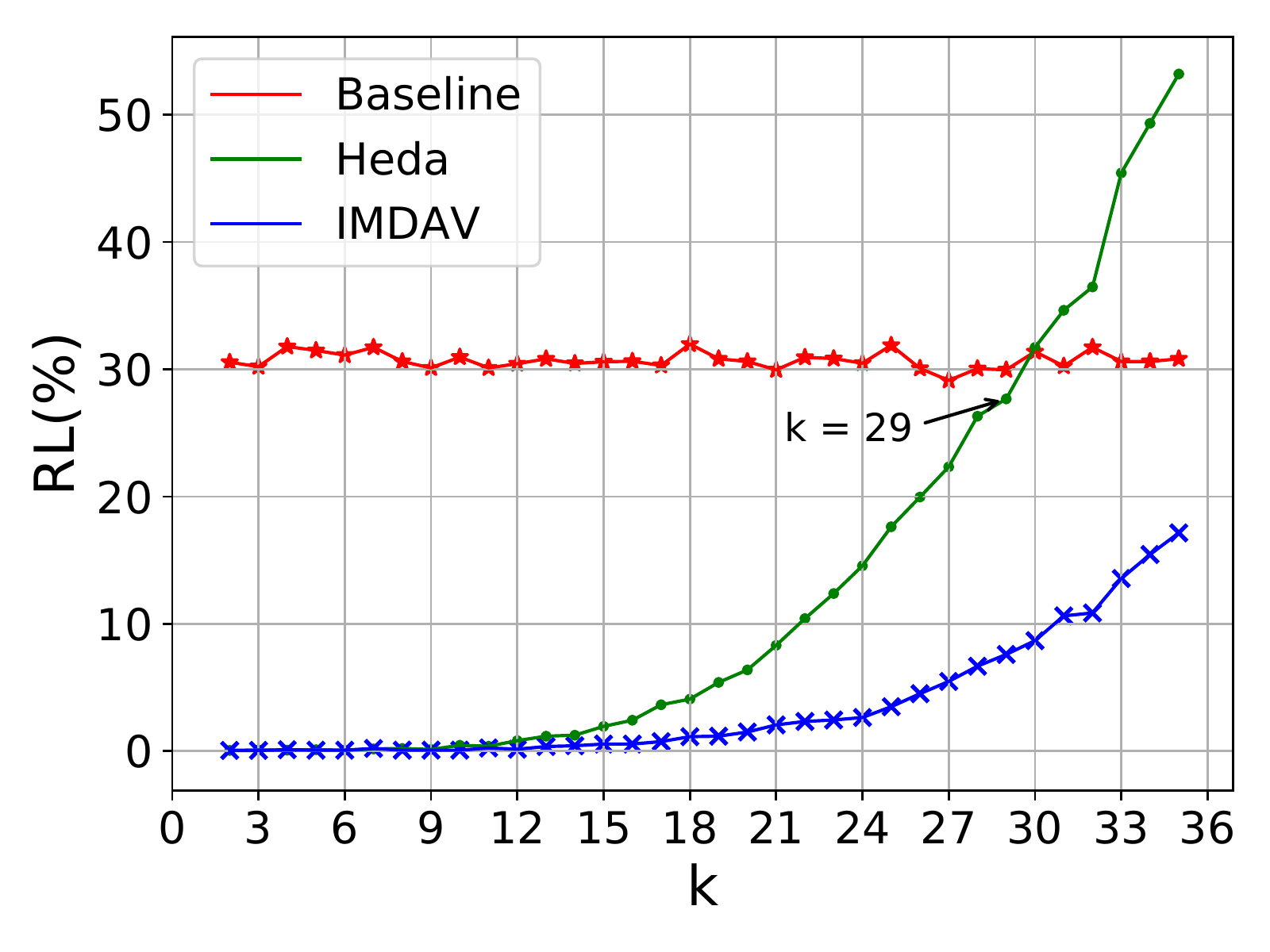}\\
    \scriptsize{(d) Car}
\end{minipage}%
\caption{Differential Privacy Components Performance: Record Linkages.}
\label{fig:Privacy Differential Components Performance: Record Linkages}
\end{figure*}

Insufficient noise may cause privacy cannot be guaranteed.
%We use RL to evaluate the risk of privacy leak.
RL is the percentage of records of the original dataset $D$ that can be correctly matched from the $\epsilon $-DP dataset ${{D}_{\epsilon }}$.
$R$ is the set of original records that are at minimum distance from $D$, and RL is defined as:
$RL=\frac{\mathop{\sum }_{{{x}_{i}}\in D}\text{Pr}\left( x_{i}^{'} \right)}{n},\ \text{where}\Pr \left( x_{i}^{'} \right)=\left\{ \begin{matrix}
   0\ if\ {{x}_{i}}\in R  \\
   \frac{1}{\left| R \right|}\ if\ {{x}_{i}}\in R  \\
\end{matrix} \right.$.
We record the RL on the four datasets for different cluster sizes and show the results in Figure \ref{fig:Privacy Differential Components Performance: Record Linkages}.
We can see from the Figure \ref{fig:Privacy Differential Components Performance: Record Linkages} that our RL is roughly the same as IMDAV, though our SSE is much lower than it, and our RL is lower than the Baseline when $k =\sqrt[{}]{\frac{m}{2}}$.
The lines in Figure \ref{fig:Privacy Differential Components Performance: Record Linkages}(b) and \ref{fig:Privacy Differential Components Performance: Record Linkages}(c) are not smooth,
the phenomenon of which may be caused by the discrete attributes in Dataset Adult and Dataset CAD.

The smaller SSE means the less loss of information,
and the smaller RL means the less risk of privacy leak.
In conjunction with Figure \ref{fig:Privacy Differential Components Performance: Sum of Squared Errors} and Figure \ref{fig:Privacy Differential Components Performance: Record Linkages}, comparing to the Baseline and IMDAV,
our DP components in \texttt{Heda} have higher security and less information loss when the cluster size is given the best value $k=\sqrt[{}]{\frac{m}{2}}$.

%, which is more obvious when the amount of data is larger (Figure \ref{fig:Privacy Differential Components Performance: Sum of Squared Errors}(a) and Figure \ref{fig:Privacy Differential Components Performance: Record Linkages}(a)).

\subsection{Evaluation of Building Blocks}\label{sec:PE-Building Blocks Performance}
The way to use our building blocks constructing a secure ML training algorithm (cf. Section \ref{sec:combine-Constructing a Specific Training Algorithms using Building Blocks}) and the security analysis of our building blocks (cf. Section \ref{sec:Security Analysis}) have been given.
In this subsection, we evaluate building blocks in terms of time consumption, accuracy and the number of interactions (round trips), taking Algorithm \ref{algorithm:Training Private-Preserving LR} as an example.
A widely used criterion -- accuracy ($\frac{\#correctly\ classified\ instances}{\#total\ instances}$) is employed to evaluate the accuracy,
and standard LR\footnote{A LR model that are trained and tested non-privately using scikit-learn (http://scikit-learn.org).} is implemented as a control group.
Table \ref{table:Building Blocks Evaluation}(a) gives the running time of each building block (Protocol \ref{protocol:Secure Addition Protocol} to Protocol \ref{protocol:Converting Ciphertext 2}) with encrypted four datasets on Algorithm \ref{algorithm:Training Private-Preserving LR},
where Protocol \ref{protocol:Converting Ciphertext 1} named Exchange 1, and Protocol \ref{protocol:Converting Ciphertext 2} named Exchange 2.
Table \ref{table:Building Blocks Evaluation}(b) gives the total time consumption, accuracy and communication overhead.

As the performance results in Table \ref{table:Building Blocks Evaluation}(a) and Table \ref{table:Building Blocks Evaluation}(b),
training a model by Algorithm \ref{algorithm:Training Private-Preserving LR} not only has almost no loss of accuracy, but also has the acceptable time consumption.
In experiments, we linearly simulate several $\mathcal{P}$, and the time consumption of $\mathcal{P}$ in Table \ref{table:Building Blocks Evaluation}(b) is the accumulation of time spent by several $\mathcal{P}$.
In actual application scenarios, several $\mathcal{P}$ conduct algorithms in parallel,
so that the time consumption of $\mathcal{P}$ and the total time consumption can be decreased sharply.
%In particular, to improve the performances, we can run several instances of building blocks in parallel.
We just show the raw running time to better illustrate the performance of building blocks.
We believe our building blocks to be practical for sensitive applications.

When the datasets become very large, the trained models have almost no loss of accuracy as shown in Table \ref{table:Building Blocks Evaluation}(b).
Results even show an increase in accuracy when evaluating on dataset Car.
The increase is because local optimal values are chosen when initializing model parameters $\beta$,
and it also shows secure training algorithms constructed by our building blocks would not affect the accuracy.
It is worth mentioning that our building blocks construct a secure LR training algorithm without use the Authorization Server and any approximate equation and it is the first solving the sigmoid function in secure LR based on HC.
%What's more,
%after combining building blocks with DP in \texttt{Heda}, efficiency is greatly improved, we evaluate  in the next subsection.
%We use DP to improve the efficiency of HC based secure training algorithm in \texttt{Heda}.
%, which is the focus of \texttt{Heda}.

\begin{table*}[!t]
\normalsize
\centering
\small
\caption{Building Blocks Performance}\label{table:Building Blocks Evaluation}
\renewcommand{\arraystretch}{1.05}
\begin{tabular}{c}
\small{(a) The running time of each building block}
\end{tabular}
\begin{tabular}{|c|c|c|c|c|c|c|c|}
\hline
Datasets &Add	&Subtraction	&Dot Product&	Multiplication	& Power Function & Exchange 1	& Exchange 2\\
\hline
\hline
\multicolumn{1}{|c|}{BCWD}&	1462ms	&268ms&	372ms	&2400ms	&2452ms&	61510ms&18ms\\		
\multicolumn{1}{|c|}{Adult}& 17026ms	&5710ms	&2759ms&	47012ms&	52291ms&	1112344ms&	17ms \\
\multicolumn{1}{|c|}{CAD}& 8643ms&	546ms&	1114ms&	18882ms&	20906ms&	411493ms&	16ms  \\
\multicolumn{1}{|c|}{Car}&	2752ms&	374ms&	890ms&	5800ms&	6515ms&	295934ms&	22ms\\
\hline
\end{tabular}
\\[1ex]
\begin{tabular}{c}
\small{(b) The total time, accuracy and communication overhead}
\end{tabular}
\begin{tabular}{|c|c|c|c|c|c|c|}
\hline
Datasets&Standard LR&	Secure LR&Total Time&$\mathcal{U}$ Time&	$\mathcal{P}$ Time&Interactions\\
\hline
\hline
\multicolumn{1}{|c|}{BCWD}&96.595\%&	95.422\%&	2428s&	32s&	2396s&	189\\
\multicolumn{1}{|c|}{Adult}&82.590\%&	81.578\%&	41145s&	459s&	40685s&	1500\\
\multicolumn{1}{|c|}{CAD}&85.607\%&	84.463\%&	17346s&	252s&	17094s&	1173\\
\multicolumn{1}{|c|}{Car}&72.32\%&	72.68\%&	10458s&	122s&	10335s&	1122\\
\hline
\end{tabular}
\end{table*}

\subsection{Evaluation of \texttt{Heda}}\label{sec:PE-Performance of Combining HC and DP}
\begin{figure}
\begin{minipage}[t]{0.25\textwidth}
    \centering
    \includegraphics[height=3.6cm]{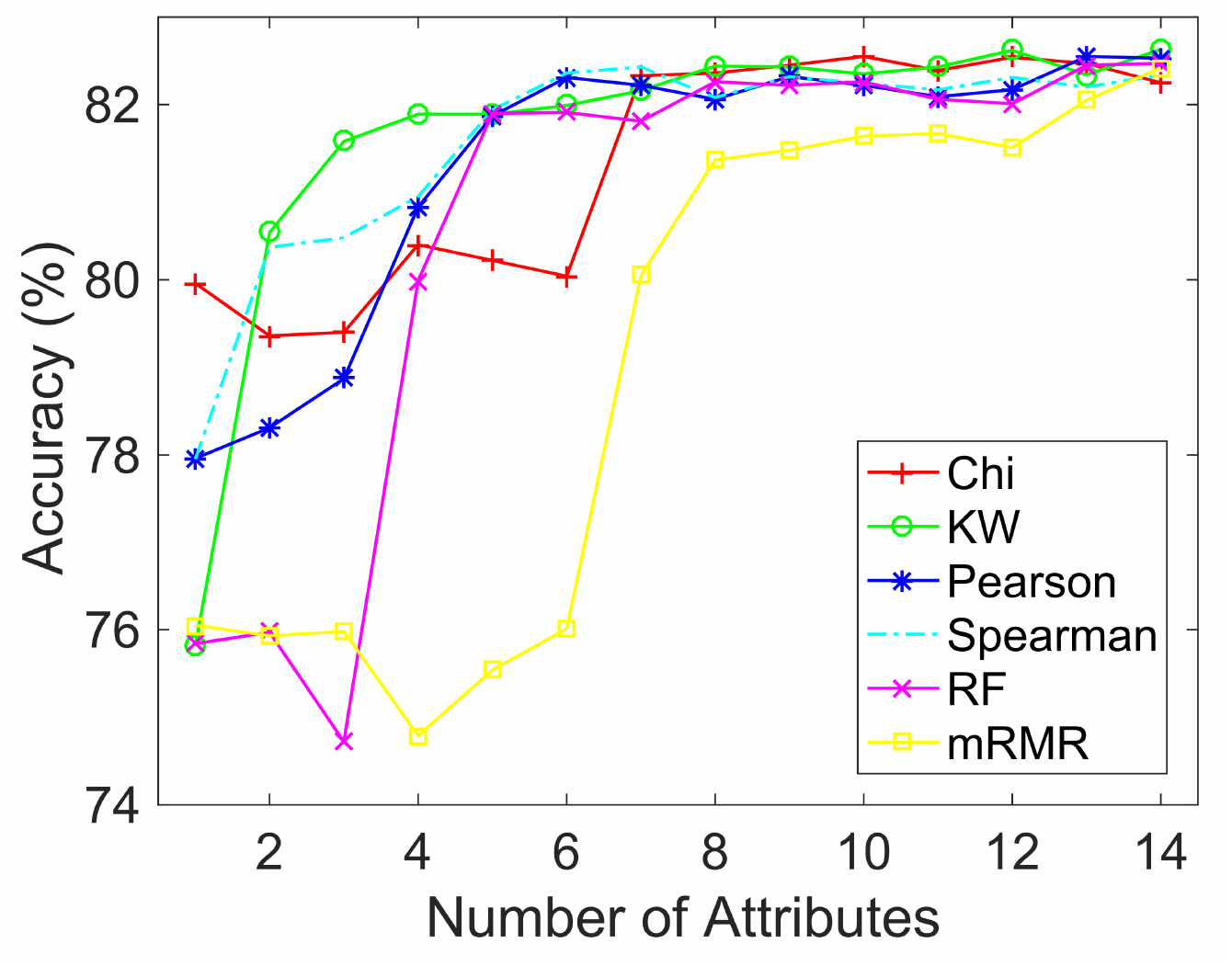}\\
    \scriptsize{(a) Adult}
\end{minipage}%
\begin{minipage}[t]{0.25\textwidth}
    \centering
    \includegraphics[height=3.6cm]{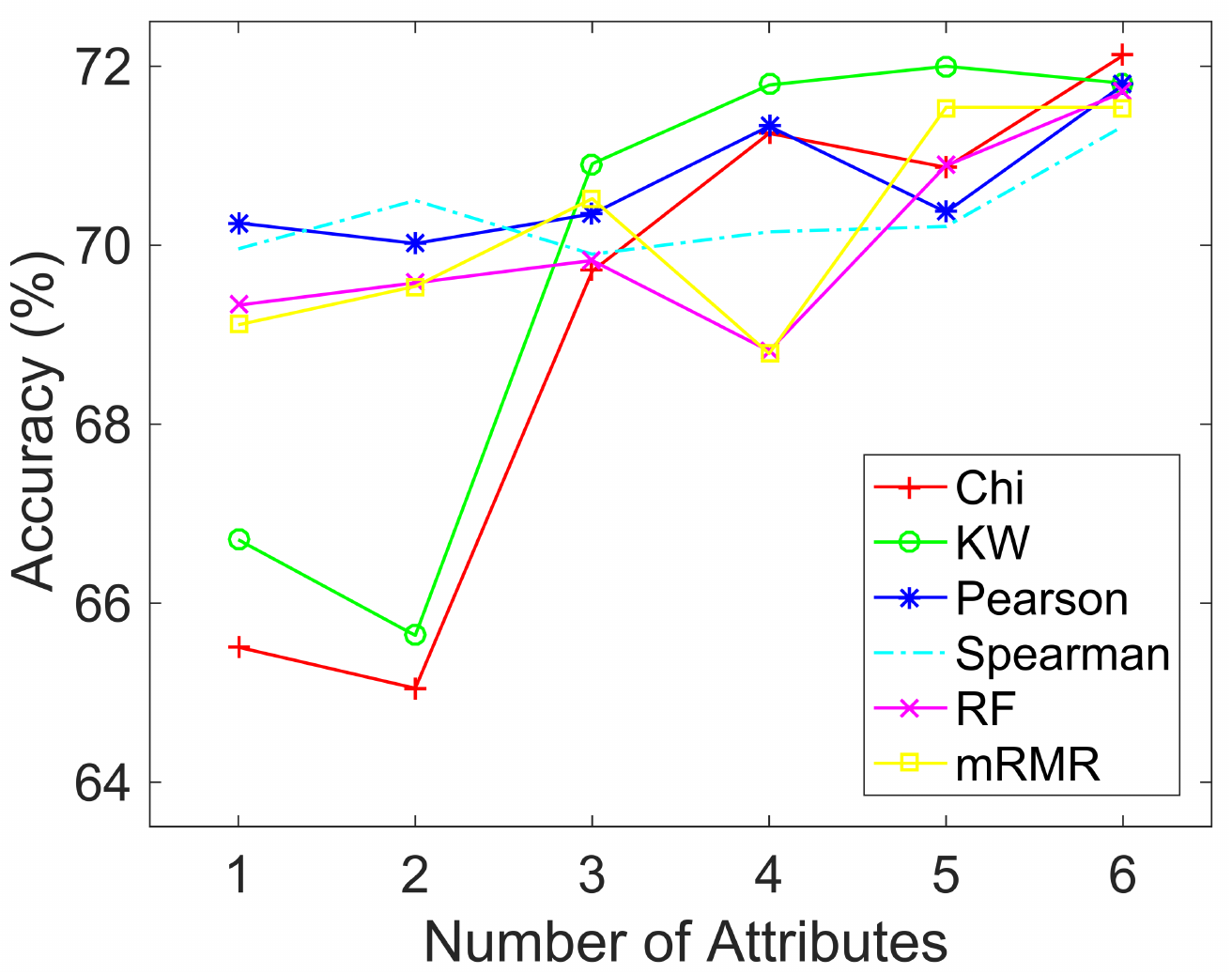}\\
    \scriptsize{(b) Car}
\end{minipage}%
\caption{Feature Evaluation Performance}
\label{fig:Feature Evaluation Proformance}
\end{figure}

We want to find a more robust feature evaluation method for serving \texttt{Heda}, i.e., finding a method that is able to divide the dataset into the high scores part and the low scores part more accurately.
We implement six widely used methods and evaluate each of them as following steps:
(i) Computing the scores of each attribute by the feature evaluation method,
(ii) According to the scores, obtaining ordered datasets (cf. Section \ref{sec:combine-Dividing a Aataset into Two Parts According to the Feature Scores}),
(iii) Obtaining a new sub-dataset by removing a feature with the lowest scores, then conducting an evaluation on this new sub-dataset,
(iv) Repeating step (iii) until there is only one dimensional data left in the dataset.
We suppose that step (iii) is able to get an ordered sub-dataset that gets the smallest decline in accuracy compared to the previous sub-dataset.
We use the standard LR evaluating each sub-dataset obtained from step (iii) and record the results.
Because of the layout restrictions, we only show the results of the dataset Adult and Car.
Adult is the dataset with a large amount of data.
Car represents the datasets that has only numerical attributes.
Figure \ref{fig:Feature Evaluation Proformance} visualizes the results, where accuracy is employed to measure the quality of datasets.

With respect to the six feature selection algorithms in Figure \ref{fig:Feature Evaluation Proformance},
we find that KW (Green lines) has a more stable performance no matter facing with a all numeric attribute dataset (Car as shown in Figure \ref{fig:Feature Evaluation Proformance}(b)) or a large amount of data dataset with numerical attributes and discrete attributes (Adult as shown in Figure \ref{fig:Feature Evaluation Proformance}(a)).
More accurate scores are able to be obtained by KW, so we use KW as our feature evaluation algorithm in the following.

\begin{figure*}
\begin{minipage}[t]{0.25\textwidth}
    \centering
    \includegraphics[height=3.5cm]{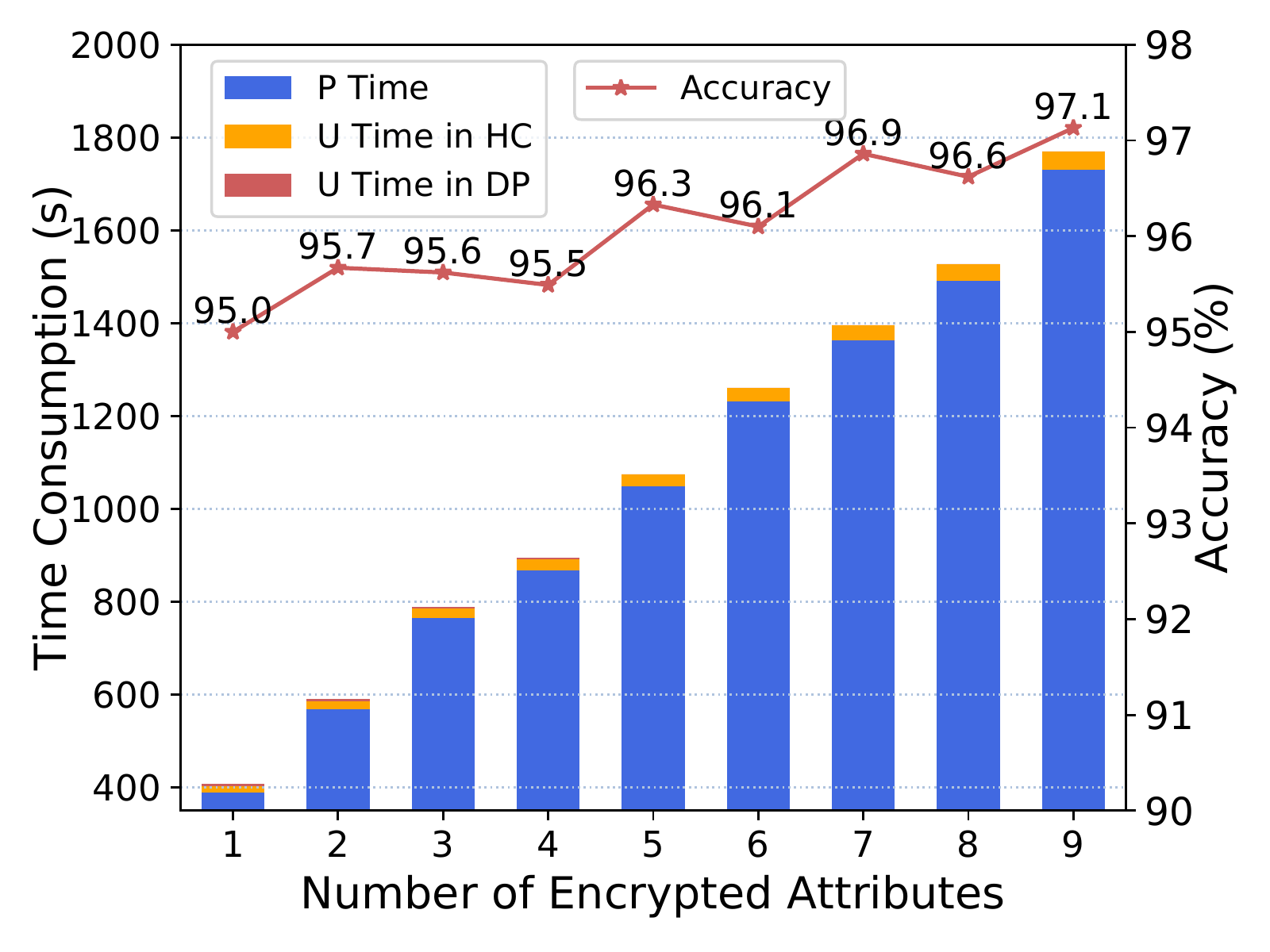}\\
    \scriptsize{(a) BCWD}
\end{minipage}%
\begin{minipage}[t]{0.25\textwidth}
    \centering
    \includegraphics[height=3.5cm]{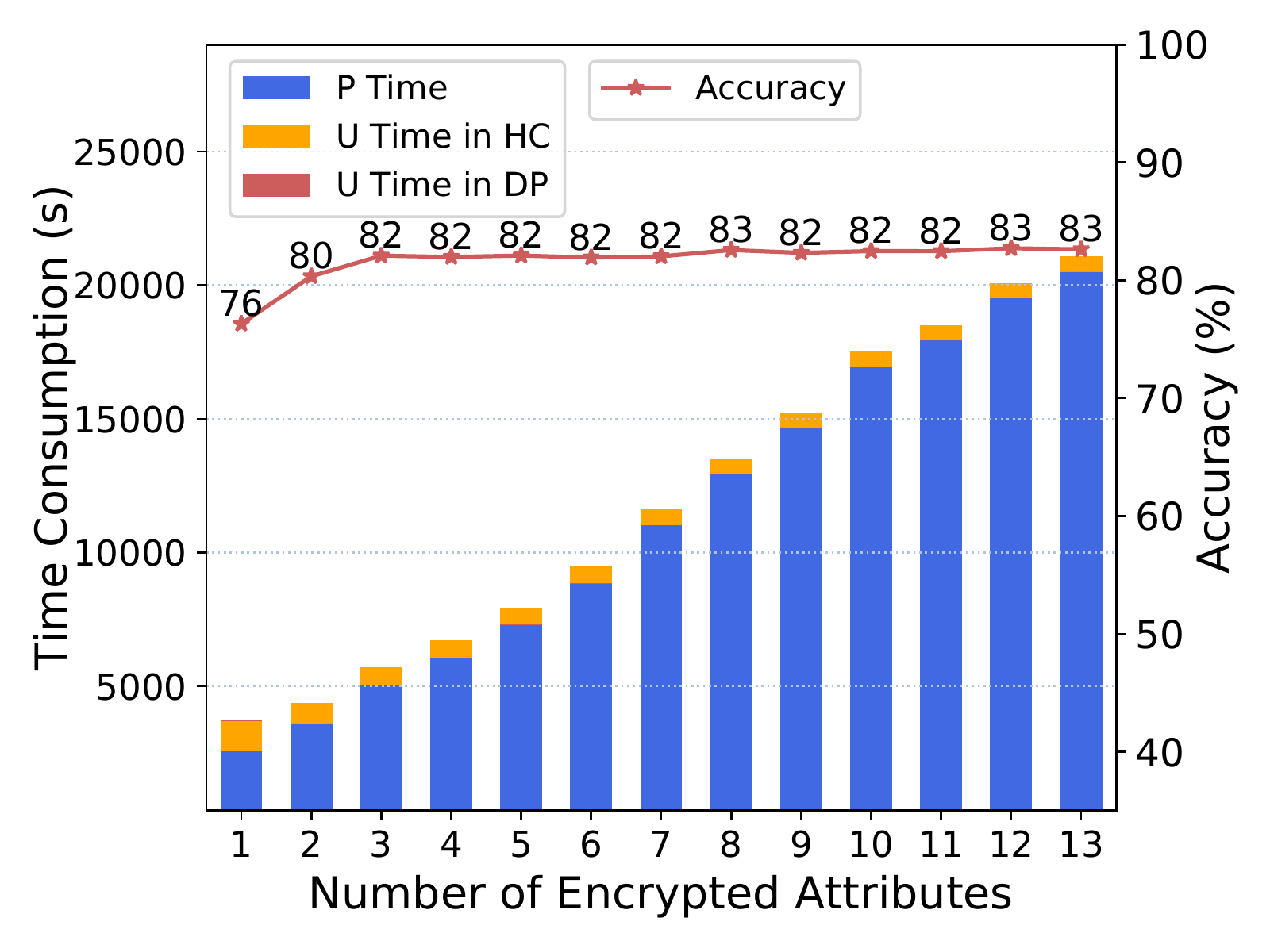}\\
    \scriptsize{(b) Adult}
\end{minipage}%
\begin{minipage}[t]{0.25\textwidth}
    \centering
    \includegraphics[height=3.5cm]{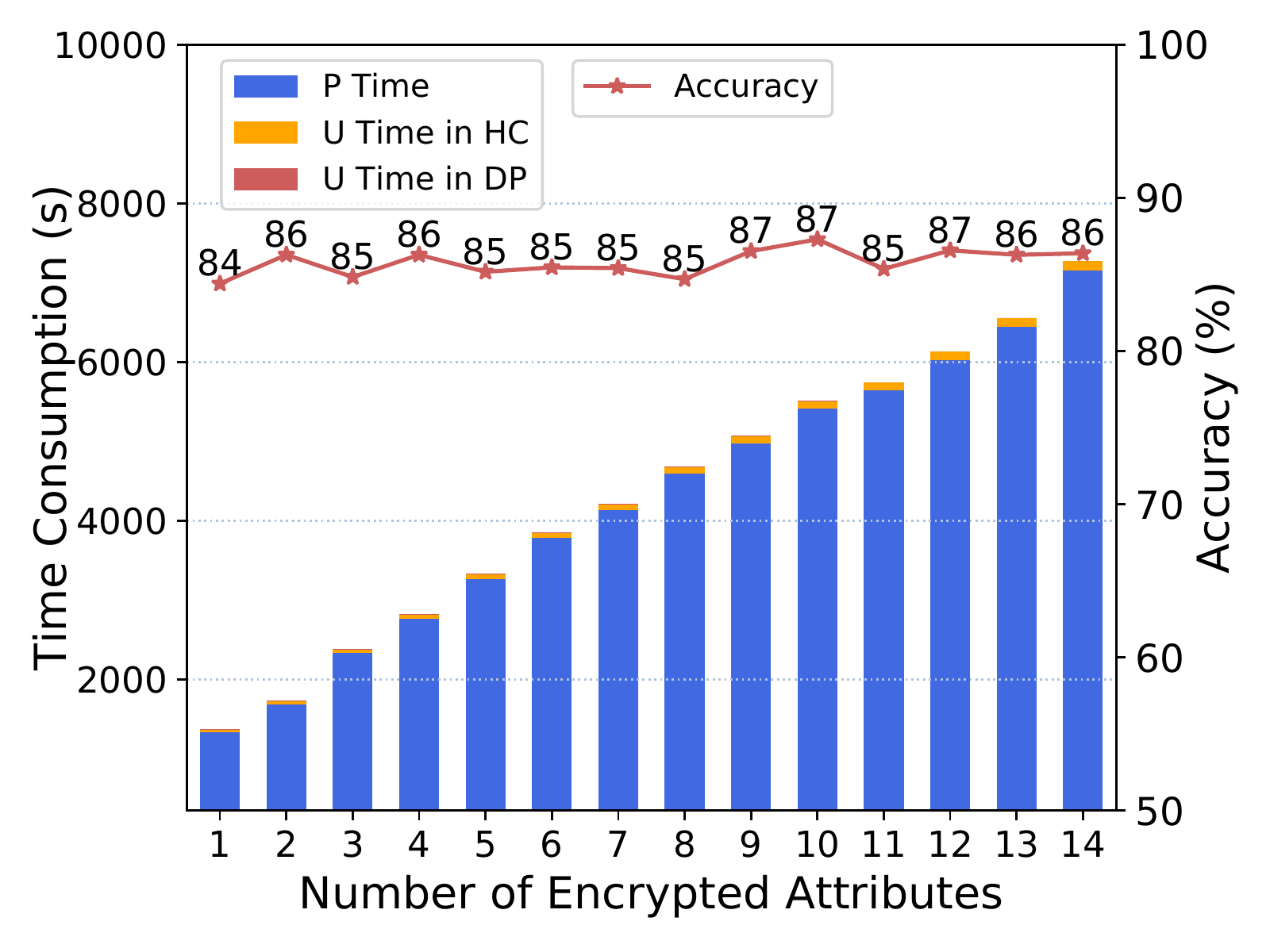}\\
    \scriptsize{(c) CAD}
\end{minipage}%
\begin{minipage}[t]{0.25\textwidth}
    \centering
    \includegraphics[height=3.5cm]{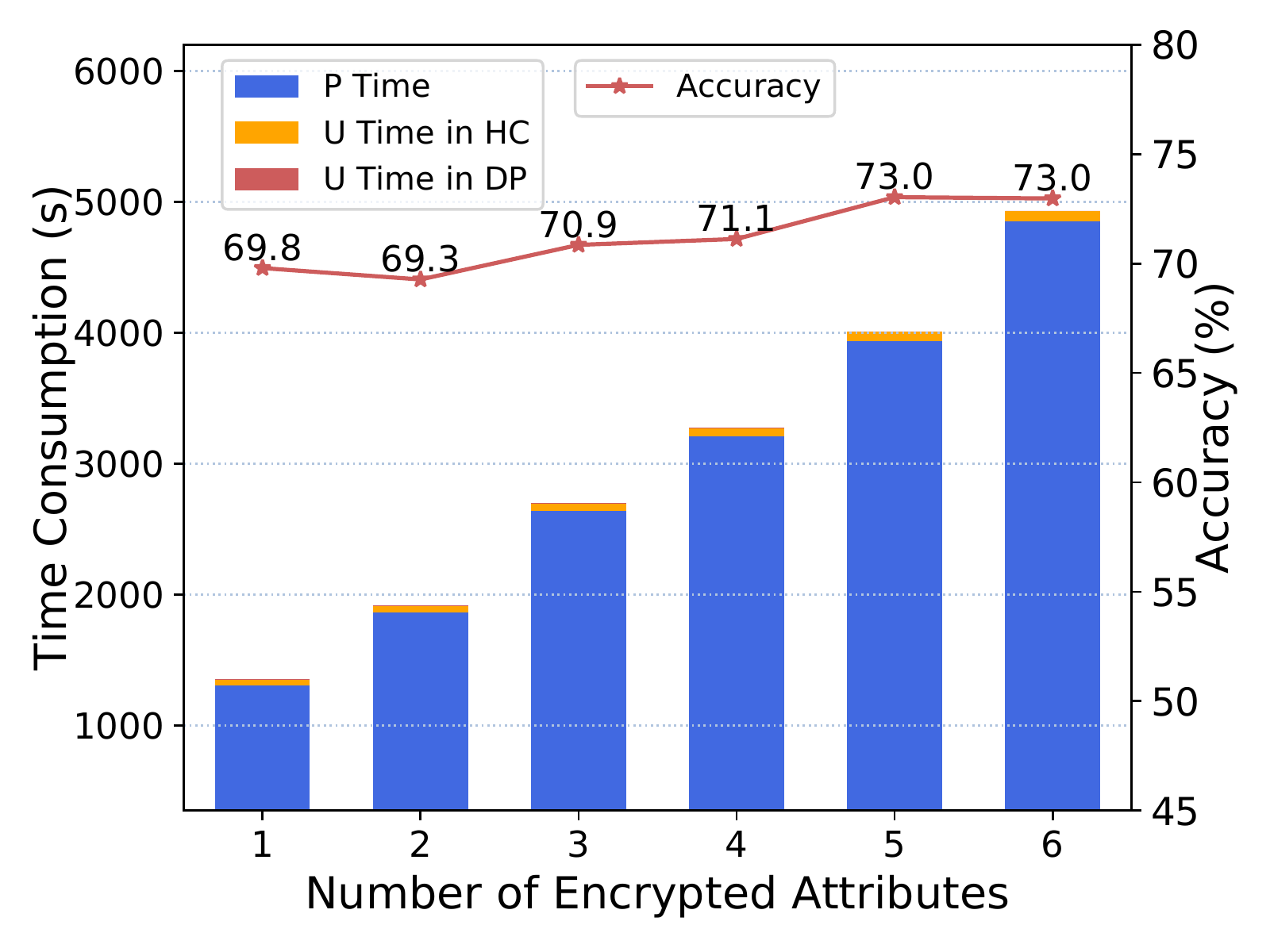}\\
    \scriptsize{(d) Car}
\end{minipage}%
\caption{\texttt{Heda} Performance. The bars represent the time consumption of \texttt{Heda}, and the line represents the accuracy.}
\label{fig:Privacy-Perserving Training evaluation}
\end{figure*}

Recalling the steps of \texttt{Heda}, after obtaining the mixed dataset, it computes model parameters on the mixed dataset.
Assuming the high scores part has $\iota$ features.
We record and observe the changes in terms of accuracy and time consumption of Algorithm \ref{algorithm:Privacy-Preserving Training} when $\iota$ changes from 1 to d.
Results are depicted in Figure \ref{fig:Privacy-Perserving Training evaluation}, where ``$\mathcal{U}$ Time in HC'' and ``$\mathcal{U}$ Time in DP'' represent the training time spent by $\mathcal{U}$ on encrypted datasets and noise datasets respectively.
After the dataset is published as the noise dataset by DP mechanism, several $\mathcal{P}$ are no longer involved in the training process, so the ``$\mathcal{P}$ Time'' is just the time consumption for several $\mathcal{P}$ participating in the encrypted dataset training.
% in Figure \ref{fig:Privacy-Perserving Training evaluation}.

\subsubsection{Efficiency}
As can be seen from Figure \ref{fig:Privacy-Perserving Training evaluation},
the total time consumption linearly increases as the $\iota$ grows, which is in line with our preset Formula \ref{equ:total time consumption}.
When $\iota=d$, \texttt{Heda} has the lowest efficiency.
But when $\iota$ changes to 1 (i.e., the number of the encrypted attributes are 1),
we obtain the highest efficiency,
and the total time consumption of \texttt{Heda} spent on each dataset can be reduced by more than 70\% compare to when $\iota=d$.
%We observe that the decrease of time consumption can be reduced by to 80\% when $\iota=1$ from Figure \ref{fig:The Increase of Time Consumption and Accuracy}.
\\\indent
When $\iota$ becomes smaller, the efficiency of \texttt{Heda} becomes higher.
The application needs to quickly train a model on sensitive datasets could set $\iota$ to one to obtain the model parameters as quickly as possible.
Results show that \texttt{Heda} is able to train a model within one hour in the face with different scales encrypted datasets.
We linearly simulate several $\mathcal{P}$, and several $\mathcal{P}$ could conduct Algorithms in parallel in practical application,
so the time consumption of $\mathcal{P}$ and the total time consumption can be decreased sharply again.
The time consumption on noise datasets is less than 1 second, because noise datasets is in the plaintext case, the speed of computing in plaintext case is much faster than computing on the encrypted data.

\subsubsection{Accuracy}
We observe from Figure \ref{fig:Privacy-Perserving Training evaluation} that training speed become faster with smaller $\iota$, but the accuracy is reduced, which indicates that the noise dataset does affect the dataset quality to a certain extent.
Nevertheless, results show that even when the minimum value of $\iota$ is taken, the trained classifier accuracy does not be reduced much (not more than 8\%).
According to the requirements of specific applications, one can adjust $\iota$ appropriately to obtain a tradeoff between accuracy and time consumption.
\\\indent
Feature evaluation techniques help ML training algorithms remove redundant or noisy attributes,
and adding appropriate noise is able to smooth out discrete attributes which has been quantized.
So the accuracy is a slightly increased, when $\iota=5$ in the mixed dataset CAD, as depicted in Figure \ref{fig:Privacy-Perserving Training evaluation}(c).

\subsubsection{Some discussions}
\texttt{Heda} obtains the tradeoffs between efficiency and accuracy by jointly applying HC and DP in an individual scheme, and enables flexible switch among different tradeoffs by parameter tuning.
A small $\iota$ is able to achieve high efficiency and a slight loss in accuracy (In our experiments, the loss is not worse than 8\%.).
Different application scenarios pay different attention to efficiency and accuracy.
According to the requirements of specific applications,
developers can obtain a balance between accuracy and efficiency by adjusting $\iota$ appropriately.

\section{Conclusion}\label{sec:conclusion}
In this paper, we proposed a novel efficient privacy-preserving ML classifier training algorithm named \texttt{Heda}.
By jointly applying HC and DP, \texttt{Heda} obtained the balance between efficiency and accuracy,
enabled flexible switch among different tradeoffs by parameter tuning.
In order to make \texttt{Heda} more efficient and accurate, we developed a library of building blocks base on HC,
gave a formula for determining the appropriate privacy budget,
and reduced the sensitivity of the query function by IMA in DP mechanism.
We demonstrated the efficiency of \texttt{Heda} and our algorithms.
%Extensive experiments demonstrated \texttt{Heda} is efficient and accurate, and
In the future work, we plan to explore a generalized framework which enables constructing a wide range of privacy-preserving ML classifier training algorithms.





%\bibliographystyle{IEEEtranS}
\label{sec:reference}
\bibliographystyle{IEEEtranS}
\bibliography{V9-bare_NDSS}

\appendix
\renewcommand\thesection{}
\subsection{Secure two-party computation}\label{sec:appendix-Secure two-party computation}
For all our two-party protocols, to ensure security,
we have to show that whatever Alice (Bob) can compute from its interactions with Bob (Alice) can be computed from its input and output,
which leads to  a commonly used definition secure two-party computation (e.g., \cite{2,9,10}).
%For more details, we refer the reader to \cite{10}.
\\\indent
We follow the notations of Bost et al. \cite{2}:
Let $F=\left( {{F}_{A}},{{F}_{B}} \right)$ be a (probabilistic) polynomial function and $\pi$ a protocol computing $F$;
Alice and Bob want to compute $F \left( a, b \right)$ where a is Alice's input and b is Bob's input, using $\pi$ and with the security parameter $\lambda$;
The view of party Alice during the execution of $\pi$ is the tuple $view_{\text{Alice}}^{\pi }(\lambda,a,b)=( \lambda ;a;{m}_{1},{{m}_{2}},...,m_{n})$ where ${m_{1}},{m_{2}},...,{{m}_{n}}$ are the messages received by Alice.
We define the view of Bob similarly.
Let $output_{Alice}^{\pi }(a,b)$ and $output_{Bob}^{\pi }(a,b)$ denote Alice's and Bob's outputs respectively.
The global output as $output^{\pi }(a,b)=(output_{Alice}^{\pi }(a,b),output_{Bob}^{\pi }(a,b))$.

\begin{myDef}\label{def:Secure Two-Party Computation}(Secure Two-Party Computation) \cite{2,10}.
A protocol $\pi $ privately computes $f$ with statistical security if for all possible inputs $(a,b)$ and simulators ${{S}_{Alice}}$ and ${{S}_{Bob}}$ hold the following properties\footnote{$\approx $ denotes computational indistinguishability against probabilistic polynomial time adversaries with negligible advantage in the security parameter $\lambda$.}:
\[\{{{S}_{Alice}},{{f}_{2}}(a,b)\}\approx \{view_{Alice}^{\pi }(a,b),outpu{{t}^{\pi }}(a,b)\}\]
\[\{{{f}_{1}}(a,b),{{S}_{Bob}}\}\approx \{outpu{{t}^{\pi }}(a,b),view_{Bob}^{\pi }(a,b)\}\]
\end{myDef}

\subsection{Modular sequential composition}\label{sec:appendix-Modular sequential composition}
For our protocols and algorithms are constructed in a modular way, we use the following useful sequential modular composition (Theorem \ref{theorem:Modular sequential composition}) \cite{15}:
%The idea of sequential modular composition (Theorem \ref{theorem:Modular sequential composition}) is that:
The Parties run a protocol $\pi$ and use calls to an ideal functionality $F$ in $\pi$ (e.g. A and B compute $F$ privately by sending their inputs to a trusted third party and receiving the results);
If we can show that $\pi$ respects privacy in the honest-but-curious model and if we have a protocol $\rho$ that privately computes $F$ in the same model, then we can replace the ideal calls for $F$ by the execution of $\rho$ in $\pi$;
the new protocol, denoted ${\pi}^{\rho}$ is then secure in the honest-but-curious model \cite{2,15}. We assume an incorruptible trusted party T for evaluating functionalities $\left( {{F}_{1}},{{F}_{2}},\ldots,{{F}_{n}} \right)-hybrid\ model$.
Parties not communicate until receiving T's output (i.e. sequential composition).
Let $\pi$ be a two-party protocol in the $\left( {{F}_{1}},{{F}_{2}},\ldots,{{F}_{n}} \right)- hybrid\ model$, and ${{\rho }_{1}},{{\rho }_{2}},\ldots,{{\rho }_{n}}$ be real protocols (i.e. protocols in the semi-honest model) computing ${{F}_{1}},{{F}_{2}},\ldots,{{F}_{n}}$.
%We follow the definition of ${{\pi }^{{{\rho }_{1}},{{\rho }_{2}},\ldots,{{\rho }_{n}}}}$ of Bost et al. \cite{2}:
All ideals calls of $ \pi $ to the trusted party for $F_{i}$ is replaced by a real execution of $\rho_{i}$.

\begin{mytheorem}(Modular Sequential Composition)\label{theorem:Modular sequential composition}
restated as in \cite{2}.
Let ${{F}_{1}},{{F}_{2}},\ldots,{{F}_{n}}$ be two-party probabilistic polynomial time functionalities and ${{\rho }_{1}},{{\rho }_{2}},\ldots,{{\rho }_{n}}$ protocols that compute respectively ${{F}_{1}},{{F}_{2}},\ldots,{{F}_{n}}$ in the presence of semi-honest adversaries.
Let $G$ be a two-party probabilistic polynomial time functionality and $\pi$ a protocol that securely computes $G$ in the $\left( {{F}_{1}},{{F}_{2}},\ldots,{{F}_{n}} \right)-hybrid\ model$ in the presence of semi-honest adversaries.
Then ${{\pi }^{{{\rho }_{1}},{{\rho }_{2}},\ldots,{{\rho }_{n}}}}$ securely computes $G$ in the presence of semi-honest adversaries.
\end{mytheorem}

\end{document}